\documentclass[11pt]{article}
\usepackage[utf8]{inputenc}
\usepackage[english]{babel}
\usepackage{amssymb,amsmath}

\begin{document}

\title{Seiberg-Witten Map with Lorentz-Invariance and Gauge-Covariant Star Product}

\author{ M.~Chaichian$^{a}$\footnote{E-mail address:masud.chaichian@helsinki.fi},  \ M.\,N.~Mnatsakanova$^{b}$\footnote{E-mail address:mnatsak@theory.sinp.msu.ru},  \ and \
 M.~Oksanen$^{a}$\footnote{E-mail address:markku.oksanen@helsinki.fi}
\vspace*{3mm} \\
\it \small  $^a$Department of Physics, University of Helsinki, P.O. Box
64, Helsinki, Finland\\
\it \small $^b$Skobeltsyn Institute of Nuclear Physics, Lomonosov
Moscow State University, \\
\it \small Moscow, Russia \\}

\date{}
\maketitle

\vspace{5mm}

\begin{abstract}
We develop the Seiberg-Witten map using the gauge-covariant star product with the noncommutativity tensor $\theta^{\mu\nu}(x)$. The latter guarantees the Lorentz invariance of the theory. The usual form of this map and its other recent generalizations do not consider such a covariant star product. We construct the Seiberg-Witten map for the gauge parameter, the gauge field and the strength tensor to the first order in the noncommutativity parameter $\theta^{\mu\nu}(x)$. Prescription for the generalization of the map to higher orders is also given. Interestingly, the associativity of the covariant star product both in the first and second orders requires the same constraints, namely, on the $\theta^{\mu\nu}(x)$ and on the space-time connection. This fact suggests that the same constraints could be enough to ensure the associativity in all orders. The resulting Seiberg-Witten map applies both to the internal and space-time gauge theories. Comparisons with the Seiberg-Witten map based on other (non-covariant) star products are given and some characteristic properties are also presented. As an application, we consider the $GL(2, C)$ noncommutative gauge theory of gravitation, in which it is shown that the connection determines a space-time with symplectic structure (as proposed by Zumino et al \cite{Zumino}). This example shows that the constraints required for the associativity of the gauge-covariant star product can be satisfied. The presented $GL(2, C)$ noncommutative gauge theory of gravitation is also compared to the one (given by Chamseddine \cite{Chamsed}) with non-covariant star product.
\end{abstract}

\section{Introduction}\label{intro}

Noncommutative field theories have been intensively studied in the last years, especially due to the hope of developing a quantum theory of gravitation, or at least to give an idea of how this could be achieved \cite{DFR} - \cite{Gr-B}.
The need for such a hope is emphasized since there is no way to quantize gravity in the canonical Hamiltonian formalism due to its nonlinearity and a formal way of   using the path integral starting with the Lagrangian of the theory, but not the Hamiltonian, cannot be justified.
One important problem which has no yet a solution is the implementation of space-time symmetries in such theories. Among the obstructions for such implementation there is the following fundamental one. If we define the noncommutativity of the space-time through the commutation relation
\begin{equation} \label{cr}
[ {x}^{\mu}, {x}^{\nu}] =i \,\theta^{\mu \nu},
\end{equation}
where $\theta^{\mu \nu}$ is a constant anti-symmetric matrix, then we arrive to the violation of the Lorentz symmetry \cite{AGM}. There are different possibilities to work with the symmetries in noncommutative field theories. One of them is to accept from the beginning that Lorentz symmetry is lost in noncommutative space-time, with all the consequences that this breakdown carry on the formulation of a given field theory. Another possibility is to consider that $\theta^{\mu \nu}(x)$ depends on coordinates and preserve the Lorentz symmetry \cite{MSSW} - \cite{DF}. In this case we have to define a star product on the space-time and assure its property of associativity, considering the dependence of  $\theta^{\mu \nu}(x)$ on coordinates. Other approaches are based on the change of space-time symmetry transformations. For example, we can impose the restriction that these transformations do not change the star product \cite{CG}.

The Seiberg-Witten map \cite{SeWi} relates gauge field theories on noncommutative spacetime to commutative theories; for recent developments see \cite{Raasakka:2010ev} - \cite{Latas:2020nji}.

Having an adequate star product defined on the space-time we can develop a noncommutative gauge theory. Then, using the Seiberg-Witten map we can manage  many of the above problems. This map can be considered as a deformation of the standard commutative gauge theory or as a low energy limit of string theory with modified backgrounds \cite{SeWi}. An extension of the Seiberg-Witten map for $x$-dependent $\theta^{\mu \nu}(x)$ was provided by \cite{MSSW} relating the non-abelian noncommutative gauge fields with their commutative counterparts. It is based on the introduction of covariant coordinates ${\hat{X}}^{\mu} = x^{\mu}  + {\hat{A}}^{\mu}$. Here, ${\hat{A}}^{\mu}(x)$ is considered as a field connected with the gauge potential ${\hat{A}}_{\mu}(x)$ through relation ${\hat{A}}^{\mu} = \theta^{\mu \nu}(x)\,{\hat{A}}_{\nu}(x)$. Then, the noncommutative tensor field ${\hat{F}}^{\mu \nu} = - i\left({\left[ {\hat{X}}^{\mu}, {\hat{X}}^{\nu}\right]}_{\star} - i\,\theta^{\mu \nu}(X)\right)$  is constructed, where the star commutator is defined as $\left[\hat{X}^{\mu},\hat{X}^{\nu}\right]_{\star}=\hat{X}^{\mu}\star\hat{X}^{\nu}-\hat{X}^{\nu}\star\hat{X}^{\mu}$. The strength tensor ${\hat{F}}_{\mu \nu}$ associated to the gauge potential ${\hat{A}}_{\mu}(x)$ is defined by the relation ${\hat{F}}^{\mu \nu} =  \theta^{\mu \rho}(x)\, \theta^{\nu \sigma}(x)\,{\hat{F}}_{\rho\sigma}$.
But, it can be verified that ${\hat{F}}_{\mu \nu}$  does not transform properly \cite{Castro}, i.e.
$\delta_{{\hat{\Lambda}}}\,{\hat{F}}_{\mu \nu} \neq {\left[ \hat{\Lambda}, {\hat{F}}_{\mu \nu}\right]}_{\star}$, despite that ${\hat{F}}^{\mu \nu}$  does have the correct law:
$\delta_{{\hat{\Lambda}}}{\hat{F}}^{\mu \nu} = {\left[ \hat{\Lambda}, {\hat{F}}^{\mu \nu}\right]}_{\star}$.
Therefore, the method of covariant coordinates does not seem to be adequate, in general, to develop a noncommutative gauge theory. Even more, it is suspected that the quantity ${\hat{A}}^{\mu}(x)$  above defined do not represent the true noncommutative gauge potential.

An alternative way to develop a noncommutative gauge theory of gravitation is to endow the space-time manifold with a gauge-covariant star product \cite{Zumino, Tagl} and generalize the Seiberg-Witten map to the case when the ordinary derivative $\partial_{\mu}$ is replaced with the covariant derivative ${\nabla}_{\mu}$ and the Moyal product is changed by a gauge-covariant one.

In this work we obtain the Seiberg-Witten map for the case when a gauge-covariant star product is defined on the space-time. We distinguish between two types of background space-times:
\begin{enumerate}
 \item [i]  the space-time is a symplectic manifold with curvature and torsion \cite{Zumino} -  \cite{ChOTZ1};
 \item [ii]  the space-time is a Poisson manifold with curvature only (torsion-free) \cite{ACG}.
\end{enumerate}
 We have to mention that in the second case the covariant star product is defined only between functions. In order to apply the corresponding Seiberg-Witten map to a gauge theory we need to generalize the respective star product \cite{ACG} to differential forms. In order to realize this, we can  extend the space-time to one with curvature and torsion, i.e. to consider the case (i).

The following argument is to invoke  the  justification  for  the use of the covariant star product and, consequently to obtain the corresponding covariant Seiberg-Witten map.
A gauge theory involves differential forms like $A = A_{\mu}dx^{\mu}$  (the 1-form gauge potential),  $F = \frac{1}{2}{F}_{\mu \nu}\,dx^{\mu} \wedge  dx^{\nu}$  (the 2-form of curvature), etc. Then we need to introduce a gauge-covariant star product between such differential forms. It has been proven that the definition of a star-product between differential forms on the space-time requires the introduction of a connection with curvature and torsion on the corresponding symplectic manifold. Therefore, we need to extend the Seiberg-Witten map to the case of noncommutative space-time endowed with such a connection.

The plan of this work is as follows.

In Section 2 we give the definition of the covariant star product between two arbitrary Lie algebra-valued differential forms and present some of its more important properties.

Section 3 is devoted to the formulation of the noncommutative gauge theory by replacing the ordinary product with the gauge-covariant star product. The integral of action for the gauge field is defined by using a volume form on the symplectic space-time manifold M of the form: ${\left({\det\theta^{\mu \nu}}\right)}^{- 1/2}\,d^{4}x$ [see equation (\ref{trace})]. The noncommutative gauge transformations that leave invariant this action are written in the usual form and we suppose that they are valid for any type of star product (Moyal, covariant, with $\theta^{\mu \nu}$ a constant anti-symmetric matrix or depending on coordinates).

In Section 4 we use the covariant star product to obtain the Seiberg-Witten map for a general gauge group. We name this map the covariant Seiberg-Witten map. The basic hypothesis we use to construct this map is to start from the standard gauge equivalence relation [see equation (\ref{geqivrel})], which we suppose to be satisfied by any type of star product (covariant or not). Then in Section 5 we consider the $GL(2, C)$ group  in order to develop a noncommutative gauge theory of gravitation. The covariant Seiberg-Witten map allows us to connect the associated noncommutative gauge potential and strength tensor with their commutative counterparts.

Section 6 is devoted to the discussion of the results and some concluding remarks on the covariant Seiberg-Witten map. The possibility of applying this map to the case of any gauge group is also discussed. This needs of course the extension of the Lie algebra to its universal enveloping. Some open questions connected with the covariant Seiberg-Witten map and its applicability to the construction of a gauge theory for gravitation are also remarked.

In the Appendices A and B we verify that the first order deformations of the gauge parameter and gauge potential are indeed solutions of the standard gauge equivalence equation (relation). The other appendices contain the expressions for the noncommutative gauge fields and strength tensor in the case of $GL(2, C)$  gauge theory of gravity.

\section{Gauge-covariant star product}

We consider a noncommutative space-time $M$ endowed with the coordinates $ x^{\mu}, \, \mu = 0,  1,  2,  3$  satisfying the commutation relation
\begin{equation} \label{cr2}
[ {x}^{\mu}, {x}^{\nu}] =i \,\theta^{\mu \nu}(x),
\end{equation}
where $\theta^{\mu \nu}(x) = - \theta^{\nu \mu}(x)$  is a Poisson bivector \cite{Zumino} - \cite{ChOTZ1}. The space-time $M$ is organized as a Poisson manifold by introducing the Poisson bracket between two functions $f(x)$ and $g(x)$ by
\begin{equation} \label{Poisbr}
\{ f, g \}_P =  \theta^{\mu \nu}\partial_\mu f \partial_\nu g.
\end{equation}
In order that the Poisson bracket satisfies the Jacobi identity, the bivector $\theta^{\mu \nu}(x)$  must obey the condition \cite{Zumino},  \cite{Tagl}, \cite{ChuHo}, \cite{ChTZ1}
\begin{equation} \label{biveccond}
 \theta^{\mu \rho} \partial_\rho  \theta^{\nu \sigma} +  \theta^{\nu \rho}  \partial_\rho   \theta^{\sigma \mu}  +   \theta^{\sigma \rho} \partial_\rho    \theta^{\mu \nu}=0.
\end{equation}
If a Poisson bracket $\{ , \}_P$ is defined on $M$, then $M$ is called a Poisson manifold (see \cite{ChuHo} for mathematical details).

Suppose now that the bivector $\theta^{\mu \nu}(x)$  has an inverse $\omega_{\mu \nu}(x)$, i.e.
\begin{equation} \label{inverse}
 \theta^{\mu \rho} \omega_{ \rho \nu} = \delta_\nu^\mu.
\end{equation}
If  $\omega = \frac{1}{2}\, \omega_{\mu \nu}\,d x^\mu  \wedge d x^{\nu}$  is nondegenerate ($\det\,\omega_{\mu \nu} \neq 0$)  and closed ($d\omega = 0$), then it is called a symplectic 2-form and $M$ is a symplectic manifold. It can be verified that the condition $d\omega = 0$ is equivalent with the equation (\ref{biveccond})  \cite{Zumino},  \cite{Tagl},  \cite{ChuHo}, \cite{HoYeh}.  In this paper we will consider only the case when $M$ is symplectic.

Because the gauge theories involve Lie algebra-valued differential forms such as $A = A_{\mu}^{\alpha}(x)\,T_{\alpha}dx^{\mu}, \: A_{\mu} = A_{\mu}^{\alpha}(x)\,T_{\alpha}$, where $T_{\alpha}$ are the infinitesimal generators of a symmetry Lie group $G$, we need to generalize the definition of the Poisson bracket to differential forms and define then an associative star product between such quantities. These problems were solved in Ref. \cite{Zumino}, \cite{Tagl}, \cite{ChOTZ1}, \cite{ChuHo},  \cite{ChTZ1},  \cite{HoYeh} and here we only report the results.	 Assuming that $\Gamma_{\mu\nu}^{\varrho} \neq \Gamma_{\nu\mu}^{\varrho}$ are the components of a nonsymmetric connection on  $M$, we can define two 1-forms of connection
\begin{equation} \label{oneforms}
\widetilde{\Gamma}_{\nu}^{\mu} = {\Gamma}_{\nu\rho}^{\mu}\,dx^{\rho}, \quad {\Gamma}_{\nu}^{\mu} = dx^{\rho} \,{\Gamma}_{\rho\nu}^{\mu}.
\end{equation}
Then we can introduce two kinds of covariant derivatives $\widetilde \nabla$  and $\nabla$, respectively. The curvatures for these two connections are respectively
\begin{eqnarray}
\widetilde R^{\nu}_{\lambda \rho \sigma} =  \partial_{\rho}\Gamma^{\nu}_{\lambda \sigma} -  \partial_{\sigma}\Gamma^{\nu}_{\lambda \rho} +
\Gamma^{\nu}_{\tau \rho}\,\Gamma^{\tau}_{\lambda \sigma} - \Gamma^{\nu}_{\tau \sigma}\,\Gamma^{\tau}_{\lambda \rho}, \label{curvatone} \\
R^{\nu}_{\lambda \rho \sigma} =	 \partial_{\rho}\,\Gamma^{\nu}_{\sigma \lambda} -  \partial_{\sigma}\,\Gamma^{\nu}_{\rho \lambda } +
\Gamma^{\nu}_{\rho \tau}\,\Gamma^{\tau}_{\sigma \lambda} -  \Gamma^{\nu}_{\sigma \tau}\,\Gamma^{\tau}_{\rho \lambda}. \label{curvat}
\end{eqnarray}
Because the connection coefficients ${\Gamma}^{\rho}_{\mu \nu}$  are not symmetric, the symplectic space-time manifold $M$  has also a torsion defined as \cite{Blago}
\begin{equation} \label{torsion}
T^{\rho}_{\mu \nu} = {\Gamma}^{\rho}_{\mu \nu} - {\Gamma}^{\rho}_{\nu \mu}.
\end{equation}
	The connection $\nabla$ satisfies the identity \cite{Zumino},   \cite{ChOTZ1}
\begin{equation} \label{nabla}
\left[ \nabla_{\mu}, \nabla_{\nu} \right]\alpha = -R^{\sigma}_{\rho \mu \nu}\,dx^{\rho}	\wedge	i_{\sigma} \alpha - T^{\rho}_{\mu \nu} \nabla_{\rho} \alpha,
\end{equation}
and an analogous formula applies for $\widetilde{\nabla}$.  Here, $\alpha$ is an arbitrary differential $k$-form, and $i_{\sigma} \alpha $  denotes the interior product which maps $\alpha$  into a $(k-1)$-form (see \cite{Zumino} for definition).

It has been proven that in order for the Poisson bracket $\{, \}_P$ to satisfy the Leibniz rule
\begin{equation} \label{Poisson}
d\,\{f, g\} = \{df, g\}	+  \{f, dg\},
\end{equation}
the bivector  $\theta^{\mu \nu}(x)$  has to obey the property \cite{Zumino} - \cite{ChOTZ1}
\begin{equation} \label{bivector}
\widetilde{\nabla}_{\rho}\theta^{\mu \nu} =   \partial_{\rho}\theta^{\mu \nu} +
 \Gamma^{\mu}_{\sigma\rho}\,\theta^{\sigma \nu} + \Gamma^{\nu}_{\sigma \rho}\,\theta^{\mu \sigma} = 0.
\end{equation}
Thus $\theta^{\mu \nu}$ is covariant constant under $\widetilde{\nabla}$, and  $\widetilde{\nabla}$ is an almost symplectic connection. If in addition to $\widetilde{\nabla}_{\rho}\theta^{\nu \nu} =  0$, one imposes ${\nabla}_{\rho}\theta^{\mu \nu} =  0$, then the torsion vanishes, $T^{\rho}_{\mu \nu} = 0$, and there is only one covariant derivative $\widetilde{\nabla} =  {\nabla}$.  As a consequence, it will not be possible to construct a gauge-covariant (associative) star product in such a case. In this paper, we do not require that ${\nabla}_{\rho}\theta^{\nu \nu}=0$.

Now, using the graded product rule for $\{, \}_{P}$, one arrives at the following general expression of the Poisson bracket between differential forms \cite{Zumino},  \cite{Tagl}, \cite{HoYeh}
\begin{equation} \label{poisbrack}
\{\alpha, \beta\}_{P} = \theta^{\rho \sigma}\,{\nabla}_{\rho}\alpha \wedge  {\nabla}_{\sigma}\beta + {(- 1)}^{|\alpha|}\, \widetilde{\mathcal{R}}^{\rho \sigma}  \wedge \left( i_{\rho}\alpha \right)  \wedge  \left( i_{\sigma}\beta \right),
\end{equation}
where $|\alpha|$ is the degree of the differential form $\alpha$, and
\begin{equation} \label{arrosig}
\widetilde{{\mathcal{R}}}^{\rho \sigma} = \frac{1}{2}\,\widetilde{{\mathcal{R}}}_{\mu \nu}^{\rho \sigma}\,dx^{\mu} \wedge  dx^{\nu}, \quad \widetilde{{\mathcal{R}}}^{\rho \sigma}_{\mu \nu} =  \theta^{\rho \lambda}\,\widetilde{{\mathcal{R}}}^{\sigma}_{\lambda \mu \nu}.
\end{equation}
In order that (\ref{poisbrack}) satisfies the graded Jacobi identity (see \cite{Zumino}, \cite{ChOTZ1}, \cite{HoYeh}, \cite{ChOTZ2} for details), the connection  ${\Gamma}^{\rho}_{\mu \nu}$  must satisfy the following additional conditions \cite{Zumino}
\begin{eqnarray} 
&&{\mathcal{R}}_{\lambda \rho \sigma}^{\nu} = 0,    \label{gammanu} \\
&&{\nabla}_{\lambda}\widetilde{{\mathcal{R}}}^{\mu \nu}_{\rho \sigma} = 0.   \label{gammamunu}
\end{eqnarray}
Having all these properties established, we can define a gauge-covariant star
product between arbitrary differential forms. Its general expression is
\begin{equation} \label{covarprod}
\alpha \star \beta = \alpha \wedge \beta + \sum\limits_{n = 1}^{\infty}\,{\left(\frac{i}{2}\right)}^{n}\,C_{n}(\alpha, \beta),
\end{equation}
where $C_{n}(\alpha, \beta)$ are bilinear differential operators satisfying the generalized Moyal
symmetry \cite{Zumino}, \cite{ChOTZ1}, \cite{HoYeh}, \cite{ChOTZ2}
\begin{equation} \label{tseen}
C_{n}\left({\alpha}^{a}, {\beta}^{b}\right) = (- 1)^{|\alpha|\,|\beta| + n}
\,C_{n}\left({\beta}^{b},{\alpha}^{a}\right).
\end{equation}
The operator $C_{1}\left({\alpha}, {\beta}\right)$  coincides with the Poisson bracket defined in equations (\ref{poisbrack}) and (\ref{arrosig}),
i.e. $C_{1}\left({\alpha}, {\beta}\right) ={ \{\alpha, \beta\}}_{P}$. An expression for  $C_{2}\left({\alpha}, {\beta}\right)$ has been also obtained in \cite{Zumino} so that the star product (\ref{covarprod}) satisfies the property of associativity
\begin{equation} \label{associat}
(\alpha \star \beta) \star \gamma = \alpha \star (\beta \star \gamma).
\end{equation}
Namely, $C_{2}\left({\alpha}, {\beta}\right)$  has the following general form
\begin{multline} \label{tsetwo}
C_{2}\left({\alpha}, {\beta}\right) = \frac {1}{2}\,\theta^{\mu \nu}\,\theta^{\rho \sigma}\nabla_{\mu}\nabla_{\rho}\alpha \wedge \nabla_{\nu}\nabla_{\sigma}\beta
{\hphantom{leftalphaleftimuirhoalpharight}}  \\
+  \frac {1}{3}\,\left(\theta^{\rho \sigma}\,\partial_{\sigma}{\theta}^{\mu \nu} +  \frac {1}{2}\,{\theta}^{\mu \sigma}\,\theta^{\nu \lambda}\,T^{\rho}_{\sigma \lambda}\right)
\left({\nabla}_{\rho}{\nabla}_{\mu}\alpha \wedge {\nabla}_{\nu}\beta - {\nabla}_{\mu}\alpha \wedge  {\nabla}_{\rho}{\nabla}_{\nu}\beta\right) \\
 - \frac {1}{2}\,\widetilde{{\mathcal{R}}}^{\mu \nu}  \wedge \widetilde{{\mathcal{R}}}^{\rho \sigma} \wedge \left(i_{\mu}i_{\rho}\alpha\right) \wedge  \left(i_{\nu}i_{\sigma}\beta\right) \\
 - \frac {1}{3}\,\widetilde{{\mathcal{R}}}^{\mu \nu}  \wedge  \left(i_{\nu}\widetilde{{\mathcal{R}}}^{\rho \sigma}\right)
 \wedge
 \left[(- 1)^{|\alpha|}\,\left(i_{\mu}i_{\rho}\alpha\right) \wedge  \left(i_{\sigma}\beta\right)
 +  \left(i_{\rho}\alpha\right) \wedge  \left(i_{\mu} i_{\sigma}\beta \right) \right] \\
 + (- 1)^{|\alpha|}\,\theta^{\mu \nu} \widetilde{{\mathcal{R}}}^{\rho \sigma} \wedge  \left(i_{\rho}{\nabla}_{\mu}\alpha\right)  \wedge  \left(i_{\sigma} {\nabla}_{\nu}\beta\right).
\end{multline}

A generalization of these results to the case of Lie algebra-valued differential forms was
done in \cite{ChOTZ1},  \cite{ChTZ1}.

Now, taking into account the graded structure of our Poisson algebra we define
the $\star$-commutator of two Lie algebra-valued differential forms $\alpha = {\alpha}^{a}\,T_{a}$
and  $\beta = {\beta}^{b}\,T_{b}$ by
\begin{equation} \label{Poisstruc}
[\alpha, \beta]_{\star} = \alpha \star \beta - {(-1)}^{|\alpha| |\beta|}\,\beta \star \alpha.
\end{equation}
For example, if $\alpha$ and $\beta$ are Lie algebra-valued one-forms, we have
\begin{equation} \label{PoisstrucLie}
[\alpha, \beta]_{\star} = {\alpha}^{a} \wedge {\beta}^{b}\,[T_{a},T_{b}]  + {\frac{i}{2}}C_{1}\left({\alpha}^a, {\beta}^b\right)\{T_{a}, T_{b}\} +  \left({\frac{i}{2}}\right)^{2}C_{2}\left({\alpha}^{a}, {\beta}^{b}\right) +{\cal O}\left({\theta}^{3}\right).
\end{equation}
This result shows that the $\star$-commutator of Lie algebra-valued differential forms does not closes, in general, in the Lie algebra but in its universal enveloping algebra. Exceptions are the unitary or some general linear complex groups where the Lie algebra closes both under the commutator $[T_{a},T_{b}]$  and the anticommutator $\{T_{a}, T_{b}\}$ of the infinitesimal generators. As an example, we will use the $GL(2, C)$ group in order to develop a noncommutative gauge theory of gravitation. The covariant Seiberg-Witten map, obtained in Section 4, will allow us to connect the associated noncommutative gauge potentials and strength tensors with their corresponding commutative quantities.

\section{Noncommutative gauge theory}
A non-Abelian gauge theory assumes a Lie algebra of a local symmetry group, whose equations of structure are
\begin{equation} \label{tab}
[T_{a},T_{b}]  = i\,f^{c}_{ab}\,T_{c}, \qquad a, b, c = 1, 2, \ldots, m.
\end{equation}
In the usual (commutative) formulation of the theory, the gauge transformations are defined as
\begin{equation} \label{deltaalfa}
\delta_{\alpha} =  i\,\alpha(x) = \alpha^{a}(x)\,T_{a}.
\end{equation}
It follows from (3.1) and (3.2) that
\begin{equation} \label{deltaalfabeta}
\delta_{\alpha}\,\delta_{\beta} - \delta_{\beta}\,\delta_{\alpha} = \delta_{-i[\alpha,\beta]}.
\end{equation}
The Lie algebra valued gauge potential $A_{\mu}(x) = A_{\mu}^{a}(x)\,T_{a}$ transforms as
\begin{equation} \label{amyu}
\delta_{\alpha}A_{\mu} = \partial_{\mu}\alpha + i\left[\alpha, A_{\mu}\right].
\end{equation}
This property allows us to define the covariant derivative and the field strength as follows
\begin{eqnarray}
&&D_{\mu} = \nabla_{\mu} - i\left[ A_{\mu}, \quad \right],	 \label{dmyu} \\
&&F_{\mu \nu}	= 	 \partial_{\mu} A_{\nu} -  \partial_{\nu} A_{\mu} - i\left[ A_{\mu},  A_{\nu}\right].	 \label{fmyunyu}
\end{eqnarray}
We remember here that the space-time $M$ is endowed with a connection $\Gamma^{\rho_{\mu \nu}}$, so that
$$
\nabla_{\mu}A_{\nu} =  \partial_{\mu} A_{\nu} - \Gamma^{\rho}_{\mu \nu} A_{\rho}.
$$
The noncommutative gauge theory is obtained by replacing the ordinary product with the covariant star product and the integral of action reads as
\begin{equation} \label{trace}
S = -{ \frac{1}{4}}Tr\int\,d^{4}x\left|P f(B)\right|\,\hat{F}^{\mu \nu} \star 	\hat{F}_{\mu \nu},
\end{equation}
where $\hat{F}_{\mu \nu}$ is the noncommutative field strength of the NC gauge potential $\hat{A}_{\mu}$ and $B = \theta^{-1}, \: \left|P f(B)\right| = \sqrt{\det B}$,  (see \cite{ChOTZ1} and \cite{HoYeh} for details). We use the hat symbol ``$\hat{\phantom{a}}$'' to denote the noncommutative quantities of our gauge theory. The action (\ref{trace}) is invariant under the noncommutative gauge transformations
\begin{equation} \label{deltaA}
\hat{\delta}_{\hat{\lambda}} \hat{A}_{\mu} = \partial_{\mu} \hat{\Lambda} + i\,{\left[ \hat{\Lambda},  \hat{A}_{\mu}\right]}_{\star}, \qquad
{\delta}_{\hat{\lambda}} \hat{F}_{\mu \nu} =  i\,{\left[ \hat{\Lambda}, \hat{F}_{\mu \nu} \right]}_{\star},
\end{equation}
where $	\hat{\Lambda}$ is the noncommutative gauge parameter ($0$-form) and
\begin{equation} \label{hatAB}
{\left[ \hat{A}, \hat{B}\right]}_{\star}	 =  \hat{A} \star \hat{B} - 	 \hat{B} \star  \hat{A}.
\end{equation}
The equation (\ref{hatAB}) applies when one of $\hat{A}$ and $\hat{B}$ is a $0$-form as is the case of (\ref{deltaA}) (see (\ref{Poisstruc}) for the general definition of the $\star$-commutator of two arbitrary differential forms). We mention also that the formulas in (\ref{deltaA}) are in accord with the gauge equivalence relation (\ref{geqivrel}) for any star product.

In what follows, we will use the gauge-covariant star product, defined in Section 2, to obtain the Seiberg-Witten map for a general gauge group. We will name this map the covariant Seiberg-Witten map.  Then in Section 5 we will use the $GL(2, C)$ group in order to develop a noncommutative gauge theory of gravitation. The covariant Seiberg-Witten map will allow us to connect the associated noncommutative gauge potential and strength tensor with the corresponding commutative quantities.

\section{Covariant Seiberg-Witten map}

The Seiberg-Witten map is defined by the gauge equivalence relation \cite{SeWi},  \cite{Ulker}
\begin{equation} \label{geqivrel}
\hat{A}_{\mu}(A; \theta) + \hat{\delta}_{\hat{\Lambda}} \hat{A}_{\mu} (A; \theta) = \hat{A}_{\mu}(A +  \delta_{\alpha} \,A; \theta),
\end{equation}
where $A$ and $\alpha$  are the ordinary (commutative) gauge field and gauge parameter, respectively, and $ \delta_{\alpha}$ is the ordinary gauge transformation given by (\ref{deltaalfa}). The equation (\ref{geqivrel}) can be rewritten as
\begin{equation} \label{geqivreltwo}
 \hat{\delta}_{\hat{\Lambda}} \hat{A}_{\mu} (A; \theta) =  \hat{A}_{\mu} (A +  \delta_{\alpha} \,A; \theta) -  \hat{A}_{\mu} (A; \theta) = 		 \delta_{\alpha} \, \hat{A}_{\mu} (A; \theta).
\end{equation}
The ordinary gauge transformation $\delta_{\alpha}$ on the right hand side of  equation (\ref{geqivreltwo}) acts on the components of $ \hat{A}$ when it is expanded as a power series in $\theta$.

Due to noncommutativity, commutators like ${\left[ \hat{\Lambda}, {\hat{A}}_{\mu}\right]}_{\star}$  take values in the enveloping algebra in the case of an arbitrary gauge group, not necessarily of the type $U(n)$. Therefore $\hat{\Lambda}$  and $ \hat{A}_{\mu}$  will also take values in this enveloping algebra. For example, for the parameter $\hat{\Lambda}$ we have \cite{Jurco}
\begin{equation} \label{hatlambda}
\hat{\Lambda} = {\Lambda}^{a}T_{a} + {\frac{1}{2}}\{T_{a}, T_{b}\}\,\Lambda^{a b} + \cdots
\end{equation}
The price we have to pay when use the enveloping algebra is that there are infinitely many parameters in the expansions of the form (\ref{hatlambda}). However, it is possible to define gauge transformations where all these infinitely many parameters depend on the usual gauge parameter $\alpha(x)$ and the gauge potential  ${A}_{\mu}(x)$ and on their derivatives. In what follows, the parameters associated to transformations of this type will be denoted by ${\hat{\Lambda}}_{\alpha}(\alpha, A; \theta)$. In other words, we restrict the enveloping algebra valued quantities to depend on their ordinary Lie algebra valued counterparts \cite{Ulker}, \cite{Jurco}. Moreover, by allowing the theory to be an enveloping algebra valued one, one can construct the NC gauge theory for an arbitrary gauge group like $SU(n)$ \cite{Jurco}.

\subsection{Case of Poisson manifold with curvature and torsion}

First, let us observe that the noncommutative gauge potential $\hat{A}_{\mu}$, the field strength $	\hat{F}_{\mu \nu}$ and the parameter $\hat{\Lambda}$ have the following functional dependence
\begin{equation} \label{funcdep}
 \hat{A}_{\mu} = \hat{A}_{\mu} (A; \theta), \quad	\hat{F}_{\mu \nu} =  \hat{F}_{\mu \nu}(A; \theta), \quad	\hat{\Lambda} = \hat{\Lambda}_{\alpha}(\alpha, A; \theta).
\end{equation}
Then it follows that we have to solve equation (\ref{geqivrel}) simultaneously for $\hat{A}_{\mu}$ and $\hat{\Lambda}$ and this is a disadvantage, specially when we like to find the higher order solutions in $\theta$ \cite{Ulker}.

This difficulty can be avoided by generalizing the ordinary gauge consistency condition (\ref{deltaalfabeta}) to the noncommutative case in the form  \cite{Ulker}, \cite{Jurco}
\begin{equation} \label{deltaalf}
i\,\delta_{\alpha}\hat{\Lambda}_{\beta} - i\,\delta_{\beta}\hat{\Lambda}_{\alpha} - {\left[\hat{\Lambda}_{\alpha},\hat{\Lambda}_{\beta}\right]}_{\star} = i\,\hat{\Lambda}_{-i[\alpha, \beta]}.
\end{equation}
Clearly, (\ref{deltaalf}) is an equation only for the gauge parameter $\hat{\Lambda}_{\alpha}$ and the solutions can be found order by order.

Let us expand the noncommutative parameter $\hat{\Lambda}_{\alpha}$ and the gauge potential $ \hat{A}_{\mu}$ as power series in $\theta^{\mu \nu}$
\begin{eqnarray}
&&\hat{\Lambda}_{\alpha} = \alpha + {\Lambda}_{\alpha}^{(1)}  + {\Lambda}_{\alpha}^{(2)} + \cdots  + {\Lambda}_{\alpha}^{(n)}  + \cdots  \; ,
\label{hatlambdalf}\\
&& \hat{A}_{\mu} = {A}_{\mu} + {A}_{\mu}^{(1)} +  {A}_{\mu}^{(2)} + \cdots  +  {A}_{\mu}^{(n)}  + \cdots  \; .
\label{hatamyu}
\end{eqnarray}
In order to obtain the gauge parameter to different orders, we introduce (\ref{hatlambdalf}) in the gauge consistency condition (\ref{deltaalf}). For the gauge-covariant star product we use expression (\ref{covarprod}) (for some generalizations see also \cite{ChOTZ2}). Applied to the calculation of products like  $\hat{\Lambda}_{\alpha} \star \hat{\Lambda}_{\beta}, \hat{\Lambda}_{\alpha} \star  \hat{A}_{\mu}$ or $ \hat{A}_{\mu}  \star  \hat{A}_{\nu}$, this star product gives the following results up to the first order in $\theta^{\mu \nu}$
\begin{eqnarray} 
&& \hat{\Lambda}_{\alpha} \star \hat{\Lambda}_{\beta} = \alpha\beta + {\frac{i}{2}}\theta^{\rho \sigma}{\nabla_{\rho}}\alpha{\nabla_{\sigma}}\beta + {\cal O}({\theta}^{2}),  \label{lambdaalbe} \\
&& \hat{\Lambda}_{\alpha} \star  \hat{A}_{\mu} = \alpha{A}_{\mu} +{\frac{i}{2}}\theta^{\rho \sigma}{\nabla_{\rho}}\alpha{\nabla_{\sigma}}{A}_{\mu}  + {\cal O}({\theta}^{2}),  \label{lambdaa}	\\
&& \hat{A}_{\mu} \star  \hat{A}_{\nu} =  {A}_{\mu}{A}_{\nu} +{\frac{i}{2}}\left(\theta^{\rho \sigma}{\nabla}_{\rho}{A}_{\mu}{\nabla_{\sigma}}{A}_{\nu} - {\frac{1}{2}}{\widetilde R}_{\mu \nu}^{\rho \sigma}{A}_{\rho}{A}_{\sigma}\right) +{\cal O}({\theta}^{2}).   \label{amuanu}
\end{eqnarray}
For different star products (Moyal, covariant, etc.), the first order term in  $\theta^{\mu \nu}$  in (\ref{lambdaalbe}), (\ref{lambdaa}) or (\ref{amuanu}) is the Poisson bracket ${\{\alpha, \beta\}}_{P},;{\{\alpha, {A}_{\mu}\}}_{P}$  or ${\{{A}_{\mu}, {A}_{\nu}\}}_{P}$  respectively, whose definition depends on the particular form of the considered star product.

Now, introducing (\ref{hatlambdalf}) in (\ref{deltaalf}) and using (\ref{lambdaalbe}) we find in the zeroth order in  the equation (\ref{deltaalfabeta}), while in the first order we obtain
\begin{equation} \label{starhatA}
i\,\delta_{\alpha}{\Lambda}_{\beta}^{(1)} - i\,\delta_{\beta}{\Lambda}_{\alpha}^{(1)}
- \left[\alpha,{\Lambda}_{\beta}^{(1)}\right] -\left[{\Lambda}_{\alpha}^{(1)},\beta\right]
- i\,\Lambda^{(1)}_{- i\,[\alpha, \beta]} =  {\frac{i}{2}}\theta^{\rho \sigma}\{{\nabla}_{\rho}\alpha, {\nabla}_{\sigma} \beta\}.
\end{equation}
Here, ${\nabla}_{\rho}\alpha = \partial_{\rho}\alpha$  and {${\nabla}_{\sigma}\beta = \partial_{\sigma}\beta$ because $\alpha$ and $\beta$  are $0$-forms (functions). In the same way, from gauge equivalence relation (\ref{geqivrel}) we deduce in the first order in $\theta$
\begin{equation} \label{geqivreltwo1st}
 \delta_{\alpha}{A}_{\mu}^{(1)} =   \partial_{\mu} {\Lambda}_{\alpha}^{(1)} + i\left[{\Lambda}_{\alpha}^{(1)}, {A}_{\mu}\right] +  i\left[{\alpha}, {A}_{\mu}^{(1)}\right]  - {\frac{1}{2}}\theta^{\rho \sigma}\{{\nabla}_{\rho}\alpha, {\nabla}_{\sigma}{A}_{\mu}\}.
\end{equation}
The solution of the equations  (\ref{starhatA}) and (\ref{geqivreltwo1st}) are respectively
\begin{equation} \label{lambdaal}
{\Lambda}_{\alpha}^{(1)} = {\frac{1}{4}}\,\theta^{\rho \sigma}\,\{{\nabla}_{\rho}\alpha, {A}_{\sigma}\},
\end{equation}
and
\begin{equation} \label{amuone}
 {A}_{\mu}^{(1)}  = - {\frac{1}{4}}\,\theta^{\rho \sigma}\{{A}_{\rho}, {\nabla}_{\sigma}{A}_{\mu} + F_{\sigma \mu}\},
\end{equation}
where $F_{\sigma \mu}$ is given by (\ref{fmyunyu}). We will verify that (\ref{lambdaal}) and (\ref{amuone}) are indeed solutions of the equations (\ref{starhatA}) and (\ref{geqivreltwo1st}) in the Appendix A.

One can find also the noncommutative field strength $\hat{F}_{\sigma \mu}$ from the definition:
$$
\hat{F}_{\mu \nu} =  \partial_{\mu} \hat{A}_{\nu} -   \partial_{\nu} \hat{A}_{\mu} - i\,{\left[  \hat{A}_{\mu}, \hat{A}_{\nu} \right]}_{\star}.
$$
In the first order in $\theta$  we obtain
\begin{equation} \label{efmunu}
F_{\mu \nu}^{(1)} =  \partial_{\mu} {A}_{\nu}^{(1)} -  \partial_{\nu} {A}_{\mu}^{(1)} -  i\left[{A}_{\mu}, {A}_{\nu}^{(1)} \right] - i\left[{A}_{\mu}^{(1)},  {A}_{\nu}\right]  -  i\left[{A}_{\mu},  {A}_{\nu}\right]_{{\star}^{1}},
\end{equation}
where ${\left[{A}_{\mu}, {A}_{\nu}\right]}_{{\star}^{1}}$ is the star commutator in the first order of $\theta$ given in (\ref{PoisstrucLie}), that is
$$
{\left[{A}_{\mu}, {A}_{\nu}\right]}_{{\star}^{1}} = \frac{i}{2}\theta^{\rho \sigma}
\left(\left\{{\nabla}_{\rho}{A}_{\mu}, {\nabla}_{\sigma}{A}_{\nu}\right\}
- {\frac{1}{2}}\,{\tilde{R}}_{\sigma \mu \nu}^{\lambda}\left\{{A}_{\rho}, {A}_{\lambda}\right\}\right).
$$
Then using (\ref{amuone}) we can write (\ref{efmunu}) in the equivalent form
\begin{equation} \label{efmunuone}
F_{\mu \nu}^{(1)} =  - {\frac{1}{4}}\,\theta^{\rho \sigma}\left(\left\{{A}_{\rho}, {\nabla}_{\sigma}F_{\mu \nu} + D_{\sigma}F_{\mu \nu}\right\}
- 2\left\{F_{\mu \rho}, F_{\nu \sigma}\right\}\right),
\end{equation}
where
\begin{equation} \label{desigefmunu}
D_{\sigma}F_{\mu \nu} = {\nabla}_{\sigma}F_{\mu \nu} - i \left[{A}_{\sigma}, F_{\mu \nu}\right]
\end{equation}
is the gauge covariant derivative of $F_{\mu \nu}$ (see the definition (\ref{dmyu})), and
\begin{equation} \label{nablasigefmunu}
{\nabla}_{\sigma}F_{\mu \nu} =   \partial_{\sigma} F_{\mu \nu} - \Gamma_{\sigma \mu}^{\rho} F_{\rho \nu}  - \Gamma_{\sigma \nu}^{\rho} F_{\mu \rho}.
\end{equation}

\subsection{Case of Poisson manifold with curvature only - torsion-free}

Suppose now that the noncommutative space-time is a Poisson manifold\linebreak
$(M,  \theta^{\mu \nu}(x))$ endowed with a torsion-free connection $\nabla$. In Ref. \cite{ACG}  a covariant star product which is associative was obtained to the third order in $\theta$. In the first order in $\theta$, this product is given by the Poisson bracket with covariant derivative. For example, if $\alpha = {\alpha}^{a}\,T_{a}$ is the Lie algebra valued gauge parameter, and $A_{\mu} = A_{\mu}^{a}\, T_{a}$ is the Lie algebra valued gauge potential, then
\begin{equation} \label{4.19}
\alpha \star {A}_{\mu} = \alpha {A}_{\mu} +{\frac{i}{2}}\,\theta^{\rho \sigma}\,{\nabla}_{\rho}\alpha\,{\nabla}_{\sigma}{A}_{\mu} +
{\cal O}({\theta}^{2}),
\end{equation}
where ${\nabla}_{\rho}\alpha \equiv \partial_{\rho}\alpha$.

As in previous case, using the gauge equivalence relation (\ref{geqivrel}), we obtain the following consistency relation for $\hat{A}_{\mu}(x)$
\begin{equation} \label{4.20}
\delta_{\alpha}{A}_{\mu}^{(1)} =  \partial_{\mu}{\Lambda}_{\alpha}^{(1)} + i\left[{\Lambda}_{\alpha}^{(1)}, {A}_{\mu}\right]
+ i\left[\alpha, {A}_{\mu}^{(1)}\right] - {\frac{1}{2}}\,{\theta}^{\rho \sigma} \left\{{\nabla}_{\rho}\alpha, {\nabla}_{\sigma}{A}_{\mu}\right\},
\end{equation}
where we used (\ref{4.19}) for the star product. This result coincides with that derived in the Section 4.1 because, in the first order in $\theta$,  the star products are identically in both cases. The gauge consistence condition for ${\Lambda}^{(1)}$  is given also by the equation (\ref{starhatA}) as in previous case of Section 4.1.

The equations (\ref{starhatA}) and (\ref{4.20}) have the solutions
\begin{equation} \label{4.21}
{\Lambda}_{\alpha}^{(1)} = {\frac{1}{4}}\,{\theta}^{\rho \sigma}\left\{{\nabla}_{\rho}\alpha, {A}_{\sigma}\right\},
\end{equation}
and respectively
\begin{equation} \label{4.22}
{A}_{\mu}^{(1)} =  - {\frac{1}{4}}\,{\theta}^{\rho \sigma}\left\{{A}_{\rho}, {\nabla}_{\sigma}{A}_{\mu} + {F}_{\sigma \mu}\right\},
\end{equation}
where
\begin{equation} \label{4.23}
{F}_{\sigma \mu} = \partial_{\sigma}{A}_{\mu} -  \partial_{\mu}{A}_{\sigma} -  i\left[{A}_{\sigma}, {A}_{\mu}\right].
\end{equation}
We verify that (\ref{4.21}) and (\ref{4.22}) are solutions of the equation (\ref{starhatA}) and respectively (\ref{4.20}) as shown in the Appendix B. It is important to observe that in this case the connection $\nabla$  is symmetric, i.e. $\Gamma_{\mu \nu}^{\rho} = \Gamma_{\nu \mu}^{\rho}$. As a consequence, we will see that (\ref{4.22}) is a solution of equation (\ref{4.20}) only if ${\nabla}_{\rho}{\theta}^{\mu \nu} = 0$, i.e. if the parameter ${\theta}^{\mu \nu}(x)$ is covariant constant. However, this does not mean that ${\theta}^{\mu \nu}(x)$ is constant, and therefore the covariant Seiberg-Witten map differs on the usual one with ${\theta}^{\mu \nu}(x) \equiv const$. This result shows again the very restrictive conditions imposed by Jacobi identity and associativity of the covariant star product.

The case presented in Section 4.2 appears more convenient to develop a gauge theory on a Riemannian noncommutative space-time (i.e. torsion-free). But we have to extend the corresponding star product \cite{ACG}  to Lie algebra valued differential forms in order to achieve the property of associativity for the star product of such quantities. In the case of Moyal star product, defined with the ordinary derivative $d_{\mu}$, this extension is not necessary because this derivative has the same form for functions, tensors or differential forms. Contrarily, the covariant derivative $\nabla_{\mu}$ differs when it acts on such quantities. Therefore, the property of associativity will not further be assured when we go from functions to the tensors in the star-product. Hawing this property satisfied, we can calculate also the higher order terms ${A}_{\mu}^{(2)}$ and ${A}_{\mu}^{(3)}$ of the noncommutative gauge field ${\hat{A}}_{\mu}$, because the star product is defined to the third order in $\theta$ in this case in Ref. \cite{ACG}.

We make now the following important observation. In both cases presented in Section 4.1 and Section 4.2, the gauge equivalence condition (\ref{geqivrel}) is satisfied, and this leads us the following general accepted transformation law of the non-commutative gauge field ${A}_{\mu}$
\begin{equation} \label{4.24}
\delta_{\hat{\Lambda}}{\hat{A}}_{\mu} = \partial_{\mu}{\hat{\Lambda}} +  i\,{\left[{\hat{\Lambda}}, {\hat{A}}_{\mu}\right]}_{\star}.
\end{equation}
This transformation law is not assured in this form when one uses covariant coordinate method (see Ref. \cite{MSSW}). Therefore, we suspect that ${\hat{A}}_{\mu}$, defined in these references by the relation ${\hat{A}}_{\mu} = {\theta}^{\mu \nu}{\hat{A}}_{\nu}$  using the covariant coordinates $\hat{X}^{\mu} = x^{\mu} + {\hat{A}}^{\mu}$, do not represent, in general, the true noncommutative gauge potential.

It is also possible to obtain general expressions for the noncommutative gauge parameter, gauge field and strength tensor in higher orders in ${\theta}^{\mu \nu}$, starting from the gauge equivalence relation (\ref{geqivrel}). The results have the same form as in the case of Moyal star product \cite{Ulker}, but they differ by the covariant star product which we are using here. Thus, in the order $(n + 1)$ in ${\theta}^{\mu \nu}$ we have
\begin{equation} \label{4.25}
\Lambda^{(n + 1)} = - {\frac{1}{4(n + 1)}}\,{\theta}^{\rho \sigma}\,\sum\limits_{p + q + r = n}{\left\{{A}_{\rho}^{(p)},  {\nabla}_{\sigma}\Lambda^{(q)}\right\}}_{\star^{r}},
\end{equation}
\begin{equation} \label{4.26}
{A}_{\mu}^{(n + 1)} = - {\frac{1}{4(n + 1)}}\,{\theta}^{\rho \sigma}\,\sum\limits_{p + q + r = n}{\left\{{A}_{\rho}^{(p)},  {\nabla}_{\sigma}{A}_{\mu}^{(q)} + {F}_{\sigma \mu}^{(q)}\right\}}_{\star^{r}},
\end{equation}
\begin{multline} \label{4.27}
{F}_{\sigma \nu}^{(n + 1)}  =
 - {\frac{1}{4(n + 1)}}{\theta}^{\rho \sigma}\sum\limits_{p + q + r = n}\left({\left\{{A}_{\rho}^{(p)},  {\nabla}_{\sigma}{F}_{\mu \nu}^{(q)} + {D}_{\sigma}{F}_{\mu \nu}^{(q)}\right\}}_{\star^{r}} \right.\\
 -\left. 2{\left\{{F}_{\mu \rho}^{(p)},  {F}_{\nu \sigma}^{(q)}\right\}}_{\star^{r}}\right),
\end{multline}
with $n =  0, 1, 2, \cdots$ and where ${\{ , \}}_{\star^{r}}$ denotes the term of $r$ order in the $\theta$ expansion of the star-anticommutator
$$
{\{\alpha, \beta\}}_{\star} = \alpha \star \beta + {(- 1)}^{|\alpha| |\beta|}\,\beta \star \alpha.
$$

As a conclusion, we emphasize again that the coordinate dependence of ${\theta}^{\mu \nu}$ appears in the expression of the covariant star product only, while the general expressions for the gauge parameter, gauge potential and strength tensor have the same general forms for any star product we use in a noncommutative gauge theory. Of course, in our case, the ordinary derivative $\partial_\mu$  is changed by covariant derivative $\nabla_{\mu}$ as we can see in equations (\ref{4.25}) - (\ref{4.27}).

\section{Noncommutative gauge theory of gravity}

As an application of the covariant Seiberg-Witten map obtained in Section 4.2, we consider the case when the symplectic space-time manifold $M$ is associated to a gauge theory of gravitation with the $GL(2, C)$ group as local symmetry (see \cite{Chamsed} -  \cite{Stern2}  for notations and definitions). The infinitesimal generators of  $GL(2, C)$  group are $(\sigma_{a b}, \gamma_{5}, I)$, where
\begin{equation} \label{5.1}
\sigma_{a b} =  - {\frac{i}{4}}\,\left[\gamma_{a}, \gamma_{b}\right],  \qquad  \gamma_{5} = i \gamma_{0} \gamma_{1} \gamma_{2} \gamma_{3}.
\end{equation}
Here, $\gamma_{a}, a = 0, 1, 2, 3$  denote the Dirac matrices and $I$ is the unit matrix.

The commutative $GL(2, C)$  gauge theory of gravitation is expressed in terms of gauge potential one-form
\begin{equation} \label{5.2}
A = A_{\mu}\,dx^{\mu}	= (\omega_{\mu} + a_{\mu}\,I + i\,b_{\mu}\,\gamma_{5})\,dx^{\mu},
\end{equation}
where $\omega_{\mu} = {\frac{i}{2}}\,\omega_{\mu}^{a b}(x)\,\sigma_{a b}$ is the spin connection, and $a_{\mu} = a_{\mu}(x)$ and $b_{\mu} = b_{\mu}(x)$ are two $U(1)$ gauge potentials \cite{Chamsed},  \cite{Stern1},  \cite{Stern2}. The corresponding $0$-form of gauge parameter is
\begin{equation} \label{5.3}
\Lambda = \lambda + \alpha\,I + i\,\beta\,\gamma_{5},
\end{equation}
where $\lambda = {\frac{1}{2}}\,\lambda^{a b}\,\sigma_{a b}$, and $\alpha$ and $\beta$ are two infinitesimal gauge parameters associated to the generators  $I$ and $\gamma_{5}$  respectively.

In addition, one introduces the $1$-form of vierbein fields
\begin{equation} \label{5.4}
E = E_{\mu}\,dx^{\mu} = \left(e_{\mu}^{a}\,\gamma_{a} + f_{\mu}^{a}\,\gamma_{5}\,\gamma_{a}\right) dx^{\mu}.
\end{equation}
The commutative  $GL(2, C)$   gauge transformations of $A$ and $E$ are
\begin{equation} \label{5.5}
\delta_{\Lambda}A = d\Lambda + i \left[\Lambda, A\right]
\end{equation}
and
\begin{equation} \label{5.6}
\delta_{\Lambda}E = i \left[\Lambda, E\right],
\end{equation}
respectively.

The $2$-forms of the curvature $F$ and torsion $T$  are
\begin{equation} \label{5.7}
F = dA - {\frac{i}{2}}\,[A, A] =  \left({\frac{1}{2}}\,R_{\mu \nu}^{a b}\,\sigma_{a b} + G_{\mu \nu}\,I + i\,H_{\mu\nu}\,\gamma_{5}\right) dx^{\mu} \wedge dx^{\nu}
\end{equation}
and
\begin{equation} \label{5.8}
T = dE - i\,[E, A] = {\frac{1}{2}}\,\left({\widetilde{T}}_{\mu \nu}^{a}\,\gamma_{a} + {\widetilde{U}}_{\mu \nu}^{a}\,\gamma_{5}\,\gamma_{a}\right) dx^{\mu} \wedge dx^{\nu},
\end{equation}
respectively. Here
\begin{equation} \label{5.9}
R_{\mu \nu}^{a b} = \partial_{\mu}\omega_{\nu}^{a b} -   \partial_{\nu}\omega_{\mu}^{a b} +  \left(\omega_{\mu}^{a c}\,\omega_{\nu}^{d b} - \omega_{\nu}^{a c}\,\omega_{\mu}^{d b}\right) \eta_{c d},
\end{equation}
\begin{align}
{\widetilde{T}}_{\mu \nu}^{a} &= {T}_{\mu \nu}^{a}  - 2 \left(f_{\mu}^{a}\,b_{\nu} -  f_{\nu}^{a}\,b_{\mu}\right), \nonumber\\
{T}_{\mu \nu}^{a} &=  \partial_{\mu}e_{\nu}^{a} -  \partial_{\nu}e_{\mu}^{a} + \left(\omega_{\mu}^{a b}\,e_{\nu}^{c} - \omega_{\nu}^{a b}\,e_{\mu}^{c}\right) \eta_{b c}, \label{5.10}
\end{align}
and
\begin{align}
{\widetilde{U}}_{\mu \nu}^{a} &= {U}_{\mu \nu}^{a}  - 2\left(e_{\mu}^{a}\,b_{\nu} -  e_{\nu}^{a}\,b_{\mu}\right), \nonumber\\
{U}_{\mu \nu}^{a} &= \partial_{\mu}\,f_{\nu}^{a} -  \partial_{\nu}f_{\mu}^{a} + \left(\omega_{\mu}^{a b}\,f_{\nu}^{c} - \omega_{\nu}^{a b}\,f_{\mu}^{c}\right) \eta_{b c}.\label{5.11}
\end{align}
In these expressions $R_{\mu \nu}^{a b}$  and ${T}_{\mu \nu}^{a}$  are the usual strength tensors connected with the curvature ${\widetilde{R}}_{\lambda \rho \sigma}^{\nu}$ and torsion ${T}_{\mu \nu}^{\rho}$  of the symplectic space-time manifold $M$ (see equations (\ref{5.23}) and (\ref{5.24})). A second torsion ${\widetilde{U}}_{\mu \nu}^{a}$ defined with the vierbein $f_{\mu}^{a}$ is obtained in the case of the $GL(2,C)$ gauge theory. Also, we denoted by ${G}_{\mu \nu}$  and ${H}_{\mu \nu}$ the strength tensors associated to the two $U(1)$ gauge potentials $a_{\mu}(x)$ and $b_{\mu}(x)$:
\begin{equation} \label{5.12}
{G}_{\mu \nu} = \partial_{\mu}a_{\nu} - \partial_{\nu}a_{\mu}, \qquad
{H}_{\mu \nu} = \partial_{\mu}b_{\nu} - \partial_{\nu}b_{\mu}.
\end{equation}

It is important to observe that it is not possible to choose the Lorentz, or equivalent $SL(2, C)$, algebra as gauge symmetry because it not longer closes upon going to the noncommutative version of the standard gauge formulation.

Let us formulate the noncommutative gauge theory of gravitation with \linebreak  $GL(2, C)$ as gauge group. The $1$-form of the gauge gravitational potential and $0$-form gauge parameter in the noncommutative case can be written, in analogy with (\ref{5.2}) and (\ref{5.3}), as
\begin{equation} \label{5.13}
\hat{A} = {\hat{A}}_{\mu}\,dx^{\mu} =  \left({\hat{\omega}}_{\mu} + {\hat{a}}_{\nu}\,I  + i\,{\hat{b}}_{\mu}\,\gamma_{5}\right) dx^{\mu}
\end{equation}
and
\begin{equation} \label{5.14}
\hat{\Lambda} = \hat{\lambda} +  \hat{\alpha}\,I  + i\,\hat{\beta}\,\gamma_{5}
\end{equation}
respectively. Remember that we use the hat symbol ``$\hat{\phantom{a}}$'' to denote the noncommutative quantities of our gauge theory. In addition, the noncommutative vierbein $1$-form is introduced in analogy with (\ref{5.4})
\begin{equation} \label{5.15}
\hat{E} = {\hat{E}}_{\mu}\,dx^{\mu} = \left({\hat{e}}_{\mu}^{a}\,\gamma_{a} + {\hat{f}}_{\mu}^{a}\,\gamma_{5}\,\gamma_{a}\right) dx^{\mu}.
\end{equation}
As a result, additional degrees of  freedom,  i.e. the gauge fields ${\hat{a}}_{\mu}$,  ${\hat{b}}_{\mu}$ and ${\hat{f}}_{\mu}^{a}$, compared to the case when  $SL(2, C)$  would be considered as gauge group, appear. However, we can use our covariant Seiberg-Witten map obtained in Section 4.1 to relate the noncommutative quantities ${\hat{A}}_{\mu}$ and $\hat{\Lambda}$ on their commutative counterparts $A_{\mu}$  and $\Lambda$. We obtain with (\ref{4.21}) and (\ref{4.22}) to the first order in $\theta$, the following expressions
\begin{equation} \label{5.16}
\hat{\Lambda} = \Lambda +  {\frac{1}{4}}\,\theta^{\rho \sigma}\left\{\nabla_{\rho}\Lambda, A_{\sigma}\right\}, \qquad \nabla_{\rho}\Lambda
\equiv \partial_{\rho}\Lambda
\end{equation}
and
\begin{equation} \label{5.17}
{\hat{A}}_{\mu} = {A}_{\mu} - {\frac{1}{4}}\,\theta^{\rho \sigma}\left\{{A}_{\rho}, \nabla_{\sigma} A_{\mu} + F_{\sigma \mu}\right\},
\end{equation}
where $F_{\sigma \mu}$ is given by (\ref{fmyunyu}). These results generalize those of  \cite{Chamsed} by using the covariant star product instead of the usual Moyal star product.

Thus the covariant Seiberg-Witten map for the vierbein  ${\hat{E}}_{\mu}$ can be obtained from the gauge equivalence condition \cite{Chamsed}
\begin{equation} \label{5.18}
{\hat{E}}_{\mu} + \delta_{\hat{\Lambda}}{\hat{E}}_{\mu} = {\hat{E}}_{\mu}\left(E + \delta_{\Lambda}E, A + \delta_{\Lambda}A\right).
\end{equation}
The solution of this equation up to the first order in $\theta^{\mu \nu}$ is
\begin{equation} \label{5.19}
{\hat{E}}_{\mu} ={E}_{\mu} - {\frac{1}{2}}\,\theta^{\rho \sigma}\left\{{A}_{\rho}, \nabla_{\sigma} E_{\mu} + {\frac{i}{2}}\,\left[{E}_{\mu},  A_{\sigma}\right]\right\}.
\end{equation}

The resulting first order expressions for the components of $1$-forms ${\hat{A}}$  and ${\hat{E}}$ are given in Appendix C.

Now we have to impose the restrictions which assure the associativity of the gauge-covariant $\star$-product. To end this, we define a nonsymmetric connection  ${\Gamma}_{\mu \nu}^{\rho}$ on the symplectic space-time manifold $M$ by
\begin{equation} \label{5.20}
{\Gamma}_{\mu \nu}^{\rho} = {\bar{e}}_{a}^{\rho}\,\left(\omega_{\nu}^{a b}\,e_{b \mu} + \partial_{\nu}e_{\mu}^{a}\right),
\end{equation}
where ${\bar{e}}_{a}^{\rho}$  is the inverse vierbein, i.e.
\begin{equation} \label{5.21}
{\bar{e}}_{a}^{\rho}\,e_{\sigma}^{a} = \delta_{\sigma}^{\rho},  \qquad  {\bar{e}}_{a}^{\rho}\,e_{\rho}^{b} = \delta_{a}^{b},
\end{equation}
and $\eta_{a b} = \mathrm{diag}(1, 1, 1, -1)$  is the Minkowski metric. The definition (\ref{5.20}) assures us that $\omega$ and $\Gamma$ represent one and the same object in two different frames on $M$ \cite{Blago}.

Because ${\Gamma}_{\mu \nu}^{\rho} \neq {\Gamma}_{\nu \mu}^{\rho}$, the symplectic space-time manifold $M$ can be endowed with a torsion having the components \cite{Blago}
\begin{equation} \label{5.22}
T_{\mu \nu}^{\rho}	= {\Gamma}_{\mu \nu}^{\rho} - {\Gamma}_{\nu \mu}^{\rho}.
\end{equation}
Also, we can define two different curvatures ${\widetilde{R}}_{\lambda \rho \sigma}^{\nu}$  (see equation (\ref{curvatone})) and $R_{\lambda \rho \sigma}^{\nu}$ (see equation (\ref{curvat})). We can prove \cite{Blago}, \cite{ZetM}  that the torsion $T_{\mu \nu}^{\rho}$  and the curvature  ${\widetilde{R}}_{\lambda \rho \sigma}^{\nu}$ associated to the non-symmetric connection (\ref{5.20}) are given, equivalently, by
\begin{equation} \label{5.23}
T_{\mu \nu}^{\rho}	= {\bar{e}}_{a}^{\rho}\,T_{\mu \nu}^{a},
\end{equation}
and respectively
\begin{equation} \label{5.24}
{\widetilde{R}}_{\lambda \rho \sigma}^{\nu} = {\bar{e}}_{\sigma}^{\nu}\,e_{b \lambda}\,R_{\rho \sigma}^{a b}.
\end{equation}

We define then the curvature tensor ${\widetilde{R}}_{\mu \nu}^{\rho \sigma}$ as in (\ref{arrosig}), and impose the restrictions (\ref{bivector}), (\ref{gammanu}) and (\ref{gammamunu}) in order to assure the associativity of the gauge-covariant star product (\ref{covarprod}) \cite{Zumino}, \cite{ChOTZ1}, \cite{ChOTZ2}. Therefore, we have to select only those $GL(2, C)$ gauge fields $\omega_{\mu}^{a b}$  and $e_{\mu}^{a}$ which satisfy the above mentioned restrictions. It is important to emphasize that we postulate that the curvature and torsion of our symplectic space-time manifold $M$ are determined only by the gauge fields   $\omega_{\mu}^{a b}$  and $e_{\mu}^{a}$, while the other gauge fields $a_{\mu}, \, b_{\mu}$  and $f_{\mu}^{a}$ are considered as auxiliary fields.

Now, we can use the expression (\ref{covarprod}) to obtain the components of the  strength tensor $\hat{F}_{\mu \nu}$, associated to the gauge potential $\hat{A}_{\mu}$, in the first order in $\theta^{\mu \nu}$. The results are presented in Appendix D.

Furthermore, the components of the noncommutative torsion $\hat{T}$ are obtained from the expression
\begin{equation} \label{5.25}
\hat{T} = d\hat{E} - i\,{\left[\hat{E}, \hat{A}\right]}_{\star} = \frac{1}{2}\,\left(T_{\mu \nu}^{a}\,\gamma_{a} + \hat{U}_{\mu \nu}^{a}\,\gamma_{5}\,\gamma_{a}\right) dx^{\mu} \wedge dx^{\nu},
\end{equation}
which is an extension of the definition (\ref{5.8}) to the noncommutative case. The results are presented in Appendix E.

Having the expression for $\hat{F}_{\mu \nu}$ and $\hat{E}_{\mu}$ we can construct an action which is gauge invariant under $GL(2,C)$ \cite{Chamsed}
\begin{multline} \label{5.26}
S = \int d^{4}x\,{\left(\det\,\theta^{\mu \nu}\right)}^{- {\frac 1 2}}\,\epsilon^{\mu \nu \rho \sigma} \mathrm{tr}\left(\left(\alpha_{1} + \beta_{1}\,\gamma_{5}\right)\,\hat{E}_{\mu} \star \hat{E}_{\nu} \star \hat{F}_{\rho \sigma} \right.\\
+\left. \left(\alpha_{2} + \beta_{2}\,\gamma_{5}\right)\,\hat{E}_{\mu} \star \hat{E}_{\nu} \star \hat{E}_{\rho} \star \hat{E}_{\sigma}\right).
\end{multline}
$\alpha^{1}, \, \beta_{1}, \, \alpha^{2}$ and $\beta_{2}$ being arbitrary constants. Here, the volume form on the symplectic space-time manifold $M$, i.e.
${\left(\det\,\theta^{\mu \nu}\right)}^{- {\frac 1 2}}\,d^{4}x$   (see equation (\ref{trace})), appears naturally and it assures the following trace property of the integral to any order in 	 $\theta^{\mu \nu}$.
\begin{equation} \label{5.27}
\int d^{4}x\,{\left(\det\,\theta^{\mu \nu}\right)}^{- {\frac 1 2}}\,\mathrm{tr}\left(\alpha \star \beta\right)
= \int d^{4}x\,{\left(\det\,\theta^{\mu \nu}\right)}^{- {\frac 1 2}}\,\mathrm{tr}\left(\beta \star \alpha\right).
\end{equation}
It can be verified that the action in equation (\ref{5.26}) is gauge invariant under $G\,L(2, C)$ because using the property (\ref{5.27}) we have
\begin{multline} \label{5.28}
\delta_{\hat{\Lambda}} S =  \int d^{4}x\,{\left(\det\,\theta^{\mu \nu}\right)}^{- {\frac 1 2}}\,\epsilon^{\mu \nu \rho \sigma}
\mathrm{tr} \left(i\,\left[\hat{\Lambda}, \left(\alpha_{1} + \beta_{1}\,\gamma_{5}\right)\hat{E}_{\mu} \star \hat{E}_{\nu} \star \hat{F}_{\rho\sigma} \right.\right.\\
+\left.\left. \left(\alpha_{2} + \beta_{2}\gamma_{5}\right)\hat{E}_{\mu}\star\hat{E}_{\nu}\star\hat{E}_{\rho}\star \hat{E}_{\sigma}\right]_{\star}\right) = 0.
\end{multline}
Using the expression (\ref{5.26}) we can obtain the field equations for the component gauge fields of ${\hat{A}}_{\mu}$ and ${\hat{E}}_{\mu}$  by imposing the principle of minimum action $\delta  S = 0$. The resulting equations would be, in general, very complicated which shows that the Seiberg-Witten map is preferable in obtaining the noncommutative gauge fields as functions of their commutative counterparts.

As a very simple example illustrating this $GL(2,C)$ noncommutative model we consider case when our symplectic space-time manifold $M$ has spherical symmetry and is associated with a teleparallel gauge theory of gravitation. Namely, we choose the gauge fields $e_{\mu}^{a}$ and $\omega_{\mu}^{a b}$ as \cite{ChOTZ1}
\begin{equation} \label{5.29}
e_{\mu}^{a} = \mathrm{diag}\left(A, 1, 1, \frac{1}{A}\right), \qquad {\bar{e}}_{\mu}^{a} = \mathrm{diag}\left(\frac{1}{A}, 1, 1, A\right), \qquad \omega_{\mu}^{a b} = 0,
\end{equation}
where $A = A(r)$  is a function depending only on the radial coordinate $r$. Then denoting the coordinates on $M$ by $(x^{\mu}) = {(r, \theta, \varphi, t)}, \: \mu = 1, 2, 3, 0$, the non-null components of the connection coefficients ${\Gamma}_{\mu \nu}^{\rho}$ defined in (\ref{5.20}) are
\begin{equation} \label{5.30}
{\Gamma}_{10}^{0} = - {\frac {A'}{A}}, \qquad {\Gamma}_{11}^{1} = \frac {A'}{A}.
\end{equation}
The only non-null components of the torsion defined in (\ref{5.22}) are
\begin{equation} \label{5.31}
T_{01}^{0} = - T_{10}^{0} = \frac {A'}{A}.
\end{equation}

The covariant derivative of the torsion tensor with the components (\ref{5.31}), defined as
\begin{equation} \label{5.32}
\nabla_{\mu}T_{\rho \sigma}^{\nu} = \partial_{\mu}T_{\rho \sigma}^{\nu} + {\Gamma}_{\mu \lambda}^{\nu}\,T_{\rho \sigma}^{\lambda} - {\Gamma}_{\mu \rho}^{\lambda}\,T_{\lambda \sigma}^{\nu} - {\Gamma}_{\mu \sigma}^{\lambda}\,T_{\rho \lambda}^{\nu},
\end{equation}
has the following non-null components
\begin{equation} \label{5.33}
\nabla_{1}T_{01}^{0} = -  \nabla_{1}T_{10}^{0} = \frac{A A'' - 2\,A'^{2}}{A^{2}}.
\end{equation}
As a result, for the two curvatures defined in (\ref{curvatone}),  (\ref{curvat}) we obtain
\begin{equation} \label{5.34}
{\widetilde{R}}_{\nu \lambda \rho}^{\mu} = 0, \qquad {\widetilde{R}}_{101}^{0} = - {\widetilde{R}}_{110}^{0} =  \frac{A A'' - 2\,A'^{2}}{A^{2}},
\end{equation}
and all other components of ${\widetilde{R}}_{\nu \lambda \rho}^{\mu}$ vanish. In these expressions, we denoted the first and second derivative of $A(r)$ by $A'$ and $A''$ respectively.

In accord with (\ref{gammanu}) we have to impose the restriction of vanishing of the curvature ${\widetilde{R}}_{\nu \lambda \rho}^{\mu}$, which is realized if the function $A(r)$  satisfies the differential equation [see (\ref{5.34})]
\begin{equation} \label{5.35}
A\,A'' - 2\,A'^{2} = 0.
\end{equation}
The solution of this equation is
\begin{equation} \label{5.36}
A(r) = - \frac{1}{\theta_{1} + \theta_{2}\,r},
\end{equation}
where $\theta_{1}$ and $\theta_{2}$ are two constants of integration. Therefore, our symplectic space-time manifold $M$ has no curvature (${\widetilde{R}}_{\nu \lambda \rho}^{\mu} = 0, \, {R}_{\nu \lambda \rho}^{\mu} = 0$) but only torsion ($T_{\mu \nu}^{\rho} \neq 0$) if we chose $A(r)$  as in (\ref{5.36}), i.e. it corresponds to a teleparallel gauge theory of gravitation. Moreover, because of the restriction (\ref{5.35}), the torsion $T_{\mu \nu}^{\rho}$ is covariant constant, $\nabla_{\lambda}T_{\mu \nu}^{\rho} = 0$, a result which is in concordance with the general property
$$
{\widetilde{R}}_{\nu \lambda \rho}^{\mu} - {R}_{\nu \lambda \rho}^{\mu} =
\nabla_{\lambda}T_{\rho \nu}^{\mu} +
\nabla_{P}T_{\nu \lambda}^{\mu} +
T_{\lambda \sigma}^{\mu}\,T_{\rho \nu}^{\sigma} +
T_{\rho\sigma}^{\mu}\,T_{\nu \lambda}^{\sigma} +
T_{\nu \sigma}^{\mu}\,T_{\lambda \rho}^{\sigma}.
$$

We chose then the noncommutative parameters $\theta^{\mu \nu}$ under the form
\begin{equation} \label{5.37}
\theta^{\mu \nu} =
\begin{pmatrix}
0 & - \frac{a}{A(r)} & - \frac{b}{A(r)} & 0 \\
\frac{a}{A(r)} & 0 & 0 & -c\,A(r) \\
\frac{b}{A(r)} & 0 & 0 & -d\,A(r) \\
0 & c\,A(r) & d\,A(r) & 0
\end{pmatrix},
\end{equation}
where $a, b, c, d$  are constants supposed to satisfy the condition $ad - bc \neq 0$, which assures the propriety of non vanishing of the $det(\theta^{\mu \nu}) = {(ad - bc)}^{2}$. It is easy to verify that parameters $\theta^{\mu \nu}$ defined in (\ref{5.37}) satisfy the restriction (\ref{bivector}).

Collecting all these results, we can obtain the expressions of the gauge fields and strength tensor to the first order in $\theta$. The results are given in the Appendix F. We can see from these results that the first order corrections to the noncommutative gauge field $\hat{\omega}_{\mu}^{a b}$ and curvature tensor  ${\hat{R}}_{\mu \nu}^{ab}$ are vanishing.

Unfortunately, all previous corrections of first order and of course of second order are very complicated even in this very simple illustrative example. They can simplify if we choose some particular expressions for the two $U(1)$  gauge fields $a_{\mu}(x), \, b_{\mu}(x)$ and the vierbeins $e_{\mu}^{a}(x), \, f_{\mu}^{a}(x)$. However, it remains open the problem of the physical signification of the two  $U(1)$   gauge fields and additional vierbein field $f_{\mu}^{a}(x)$. Maybe, this problem could be solved in a unified gauge theory for gravitation with the other fundamental interactions.

It would be very interesting to develop a noncommutative gauge theory of gravity using other gauge groups like Poincar\'{e} group. In such cases it may be necessary to extend the Lie algebra to its universal enveloping algebra. However, there is a major difficulty in defining the noncommutative curvature and torsion tensors using expressions similar with (\ref{5.23}), (\ref{5.24}). This problem remains open for new researches on the applicability of the Seiberg-Witten map to the different types of star products.

\section{Conclusions and discussions}

In our work we obtained the Seiberg-Witten map for a noncommutative gauge theory developed over the symplectic space-time manifold endowed with the covariant star-product.
In other words,  we  have deduced the covariant Seiberg-Witten map where the star-product is defined by using the covariant derivative instead of ordinary one.
The motivation for using the covariant star-product is that the gauge theory involves differential forms like  $A = A_{\mu}\,dx^{\mu}$ (the 1-form gauge potential),  $F =\frac{1}{2}F_{\mu \nu}\,dx^{\mu} \wedge dx^{\nu}$ (the 2-form of curvature ) etc.
It has been proven that the definition of a star-product between differential forms on the space-time requires the introduction of a connection with curvature and torsion on the corresponding space-time manifold. Therefore, we need to extend the Seiberg-Witten map to the case of noncommutative space-time endowed with such a connection and use a covariant-star product.

The basic idea which we adopted to construct the covariant Seiberg-Witten map was to impose the standard gauge equivalence relation (see equation (\ref{geqivrel}), which we supposed to be satisfied by any type of star-product (covariant or not). Then we obtained the first order deformations in $\theta^{\mu \nu}$ for the gauge parameter and gauge potential. It was possible also to express the higher order deformations for gauge parameter, gauge potential and strength tensor in the same form as in the standard case of a noncommutative theory. We concluded that the dependence of $\theta^{\mu \nu}(x)$  on coordinates appears in the expression of the covariant star product only, while the general expressions for the gauge parameter, gauge potential and strength tensor have the same general forms for any star product we use in a noncommutative gauge theory.

We applied our results to the case when the space-time is a symplectic manifold $M$ associated to a gauge theory of gravitation with the  as local group of symmetry. We determined the gauge parameter, gauge potential and strength tensor in the first order in noncommutatitiy parameters . Their expressions are similar with those obtained by using a Moyal star-product \cite{Ulker}, but with the ordinary derivative changed by the covariant derivative. However, in higher orders in $\theta^{\mu \nu}$  they will be different because the covariant star-product differs of Moyal product not only by the covariant derivative but even by terms which include the derivatives of the parameters $\theta^{\mu \nu}$.

As a very simple example illustrating the $GL(2,C)$  noncommutative model we considered case when the symplectic space-time manifold $M$ has spherical symmetry and is associated with a teleparallel gauge theory of gravitation. This example showed that it is possible to accommodate the gauge connection so that it determines a space-time having a symplectic structure and which is endowed with a covariant star-product. This means that it is possible to impose on the gauge connection all the restrictions required by the covariant star-product.

In order to apply the covariant Seiberg-Witten map to other gauge theories of gravitation, as would be for example those where the gauge group is Poincar\'{e}  or de-Sitter \cite{ChTSZ}, \cite{ChTZ2}, we need to extend the Lie algebra to its universal enveloping. However, there is a major difficulty in defining the noncommutative curvature and torsion tensors using expressions similar with those from the case when Lie algebra closes both under commutator and anti-commutator of the infinitesimal generators.

It would be also interesting to obtain the covariant Seiberg-Witten map in the case of a gauge theory defined on a noncommutative space-time endowed with curvature only (torsion-free). However, this will require the construction of a covariant star-product between differential forms on such a space-time. For example, we could try to extend the results of \cite{ACG}, which are valid for the star-product between functions only, to the case of differential forms. We believe that this extension will impose the introduction of a connection with curvature and torsion, i.e. we will arrive to the structure described in \cite{Zumino}. Therefore, the construction of a covariant star-product between differential forms defined over a space-time manifold having only curvature (without torsion) remains an open problem for new researches.

\appendix
\numberwithin{equation}{section}

\section{Appendix A}

Here we verify that (\ref{lambdaal}) and (\ref{amuone}) are indeed solutions of (\ref{starhatA}) and respectively (\ref{geqivreltwo1st}). Because the equation (\ref{starhatA}) is identically with that from the case when $\theta^{\mu \nu}$ is constant, i.e. the dependence of $\theta^{\mu \nu}$ on coordinates does not bring supplementary terms in (\ref{starhatA}), the solution (\ref{lambdaal}) is the same as for $\theta^{\mu \nu} = constant$.

This is not the case for the equation (\ref{geqivreltwo1st}). Indeed, using the solutions (\ref{lambdaal}) and (\ref{amuone}), we can write the equation (\ref{geqivreltwo1st}) under the form
\begin{multline}\label{a1}
\delta_{\alpha}A_{\mu}^{(1)} = \delta_{\alpha}\bar{A}_{\mu}^{(1)} + \frac{1}{4}\left(\partial_{\mu}\theta^{\rho \sigma}\right) \left\{\partial_{\rho}\alpha, A_{\sigma}\right\} \\ +  \frac{i}{4}\,\theta^{\rho \sigma}\,\Gamma_{\sigma \mu}^{\lambda}\,\left[\alpha, \left\{A_{\rho}, A_{\lambda}\right\}\right] +
\frac{1}{2}\,\theta^{\rho \sigma}\,\Gamma_{\sigma \mu}^{\lambda}\left\{\partial_{\rho}\,\alpha, A_{\lambda}\right\},
\end{multline}
where	we denoted
\begin{equation}\label{a2}
A_{\mu}^{(1)} = \bar{A}_{\mu}^{(1)} + \frac{1}{4}\,\theta^{\rho \sigma}\,\Gamma_{\sigma \mu}^{\lambda}\left\{\partial_{\rho}\alpha, A_{\lambda}\right\},
\end{equation}
and
\begin{equation}\label{a3}
 \bar{A}_{\mu}^{(1)} = - \frac{1}{4}\,\theta^{\rho \sigma}\left\{A_{\rho}, \partial_{\sigma}A_{\mu} + F_{\sigma \mu}\right\}.
\end{equation}
Now we compute $\delta_{\alpha}A_{\mu}^{(1)}$  using the solution (\ref{a2})
\begin{equation}\label{a4}
\delta_{\alpha}A_{\mu}^{(1)} = \delta_{\alpha}\bar{A}_{\mu}^{(1)} + \frac{1}{4}\,\theta^{\rho \sigma}\,\Gamma_{\sigma \mu}^{\lambda}\,\delta_{\alpha}\left\{A_{\rho}, A_{\lambda}\right\}.
\end{equation}
But
\begin{equation}\label{a5}
\delta_{\alpha}\left\{A_{\rho}, A_{\lambda}\right\} = \left\{\partial_{\rho}\alpha, A_{\lambda}\right\} +  \left\{\partial_{\lambda}\alpha, A_{\rho}\right\} + i\left[\alpha\left\{A_{\rho}, A_{\lambda}\right\}\right],
\end{equation}
and therefore we can write (\ref{a4}) as
\begin{multline}\label{a6}
\delta_{\alpha}A_{\mu}^{(1)} =  \delta_{\alpha}\bar{A}_{\mu}^{(1)} +
\frac{1}{4}\,\theta^{\rho \sigma}\,\Gamma_{\sigma \mu}^{\lambda}\, \left\{\partial_{\rho}\alpha, A_{\lambda}\right\} \\ +
\frac{1}{4}\,\theta^{\rho \sigma}\,\Gamma_{\sigma \mu}^{\lambda}\, \left\{\partial_{\lambda}\alpha, A_{\rho}\right\} +
\frac{i}{4}\,\theta^{\rho \sigma}\,\Gamma_{\sigma \mu}^{\lambda}\,\left[\alpha, \left\{A_{\rho}, A_{\lambda}\right\}\right].
\end{multline}
Identifying (\ref{a1}) with (\ref{a6}) we obtain the following condition that assures that (\ref{amuone}) is solution of the equation (\ref{geqivreltwo1st})
\begin{eqnarray}\label{a7}
\frac{1}{4}\left(\partial_{\mu}\theta^{\rho \sigma}\right)\left\{\partial_{\rho}\alpha, A_{\sigma}\right\}  +
\frac{i}{4}\,\theta^{\rho \sigma}\,\Gamma_{\sigma \mu}^{\lambda}\,\left[\alpha, \left\{A_{\rho}, A_{\lambda}\right\}\right] +
\frac{1}{2}\,\theta^{\rho \sigma}\,\Gamma_{\sigma \mu}^{\lambda}\left\{\partial_{\rho}\alpha, A_{\lambda}\right\}  \nonumber \\
= \frac{1}{4}\,\theta^{\rho \sigma}\,\Gamma_{\sigma \mu}^{\lambda}\left\{\partial_{\rho}\alpha, A_{\lambda}\right\} +
\frac{1}{4}\,\theta^{\rho \sigma}\,\Gamma_{\sigma \mu}^{\lambda}\left\{\partial_{\lambda}\alpha, A_{\rho}\right\} +
\frac{i}{4}\,\theta^{\rho \sigma}\,\Gamma_{\sigma \mu}^{\lambda}\,\left[\alpha, \left\{A_{\rho}, A_{\lambda}\right\}\right].
\end{eqnarray}
Equivalently, (\ref{a7}) can be written as
\begin{equation}\label{a8}
\left[\partial_{\mu}\theta^{\rho\lambda} + \Gamma_{\sigma \mu}^{\rho}\,\theta^{\sigma \lambda} +
\Gamma_{\sigma \mu}^{\lambda}\,\theta^{\rho \sigma}\right]\left\{\partial_{\rho}\alpha, A_{\lambda}\right\}.
\end{equation}
It is known that in order that the Poisson bracket ${\{ , \}}_{P}$ satisfies the Leibniz  rule, the bivector $\theta^{\mu \nu}(x)$ has to obey the property (\ref{bivector}), i.e.
\begin{equation}\label{a9}
{\tilde{\nabla}}_{\mu}\theta^{\rho \lambda} = \partial_{\mu}\theta^{\rho \lambda} + \Gamma_{\sigma \mu}^{\rho}\,\theta^{\sigma \lambda} +  \Gamma_{\sigma \mu}^{\lambda}\,\theta^{\rho \sigma} = 0.
\end{equation}
Then the condition (\ref{a8}) is verified and this means that (\ref{amuanu}) is indeed the solution of the equation (\ref{geqivreltwo1st}).

We remark here that on a Poisson manifold endowed with a non-symmetric connection, there is another covariant derivative ${\nabla}_{\mu}$ with the property that, in general,
\begin{equation*}
{\nabla}_{\mu}\theta^{\rho \lambda} = \partial_{\mu}\theta^{\rho \lambda} + \Gamma_{\mu \sigma}^{\rho}\,\theta^{\sigma \lambda} +  \Gamma_{\sigma \mu}^{\lambda}\,\theta^{\rho \sigma} \neq 0
\end{equation*}
(for details see Refs.\cite{Zumino}  and \cite{ChOTZ1}).

As a conclusion, we stress that because of the restrictions imposed on bivector $\theta^{\mu \nu}(x)$  (Jacobi identity and property (\ref{a9})), the derivatives of $\theta^{\mu \nu}(x)$  do not appear at first order in the Seiberg-Witten map.

\section{Appendix B}

Here we verify that (\ref{4.22}) is a solution of the equation (\ref{4.20}). The consistency condition for the gauge parameter is the same as in the case of a noncommutative Minkowski space-time. So it remains to verify the consistence condition (\ref{4.20}) for the gauge field only.

As in Appendix A, we find that (\ref{4.22}) is a solution of the equation (\ref{4.20}) if the following relation is satisfied:
\begin{equation}\label{b1}
\left[ \partial_{\mu}\theta^{\rho \lambda} +  \Gamma_{\sigma \mu}^{\rho}\,\theta^{\sigma \lambda} +  \Gamma_{\sigma \mu}^{\lambda}\,\theta^{\rho \sigma} \right]\left\{\partial_{\rho}\alpha, A_{\lambda}\right\} = 0.
\end{equation}
But in this case the connection $\nabla$ is symmetric, i.e. $\Gamma_{\mu \nu}^{\rho} = \Gamma_{\nu \mu}^{\rho}$. As a consequence, the relation (\ref{b1}) can be written as
\begin{equation}\label{b2}
\left[ \partial_{\mu}\theta^{\rho \lambda} +  \Gamma_{\mu \sigma}^{\rho}\,\theta^{\sigma \lambda} +  \Gamma_{\mu \sigma}^{\lambda}\,\theta^{\rho \sigma} \right]\left\{\partial_{\rho}\alpha, A_{\lambda}\right\} \equiv {\nabla}_{\mu}\theta^{\rho \lambda}\left\{\partial_{\rho}\alpha, A_{\lambda}\right\} = 0.
\end{equation}
Equivalently, the relation (\ref{b2}) imposes the restriction
\begin{equation}\label{b3}
{\nabla}_{\mu}\theta^{\rho \lambda} = 0.
\end{equation}
The conclusion is that in the first order in $\theta$, the Seiberg-Witten map is identically in the two cases 4.1 and 4.2, but in the case 4.2 we have to impose the condition of constant covariance for deformation parameter $\theta^{\mu \nu}(x)$. This restriction is not necessary in the case 4.1 where the connection  $\nabla$  is non-symmetric, i.e. $\Gamma_{\mu \nu}^{\rho} \neq \Gamma_{\nu \mu}^{\rho}$, and where the condition ${\tilde{\nabla}}_{\mu}\theta^{\rho \lambda} = 0$ assures that (\ref{4.22}) is a solution of the equation (\ref{4.20}) (see Refs. \cite{Zumino}, \cite{ChOTZ1}, \cite{ChTZ1} and \cite{HoYeh} for details).

\section{Appendix C}
Here we give the expressions of the noncommutative $GL(2,C)$ gauge fields to the first order in  obtained by using the covariant Seiberg-Witten map
\begin{multline}\label{c1}
{\hat{\omega}}_{\mu}^{ab} =  \frac{1}{2}\,\theta^{\rho \sigma}
\left[{\omega}_{\rho}^{ab}\left({\nabla}_{\sigma}a_{\mu} + G_{\sigma\mu}\right) + a_{\rho}\left({\nabla}_{\sigma}{\omega}_{\mu}^{ab} + R_{\sigma\mu}^{ab}\right) \right.\\
- \left. \frac{1}{2}\,\varepsilon^{ab}_{cd}{\omega}_{\rho}^{cd}\left({\nabla}_{\sigma}b_{\mu} + H_{\sigma\mu}\right) -
\frac{1}{2}\varepsilon^{ab}_{cd}\,b_{\rho}\right],
\end{multline}
\begin{multline}\label{c2}
{\hat{a}}_{\mu} = a_{\mu} - \frac{1}{2}\,\theta^{\rho \sigma}\left[a_{\rho}\left({\nabla}_{\sigma}\,a_{\mu} + G_{\sigma\mu}\right)  -
b_{\rho}\,\left({\nabla}_{\sigma}b_{\mu} + H_{\sigma\mu}\right)  \right. \\ - \left. \frac{1}{8}\,\eta_{ac}\,\eta_{bd}\,{\omega}_{\mu}^{cd}\,
\left({\nabla}_{\sigma}{\omega}_{\mu}^{ab} + R_{\sigma\mu}^{ab}\right)\right],
\end{multline}
\begin{multline}\label{c3}
{\hat{b}}_{\mu} = b_{\mu} - \frac{1}{2}\,\theta^{\rho \sigma}\left[a_{\rho}\left({\nabla}_{\sigma}b_{\mu} + H_{\sigma\mu}\right)  +
b_{\rho}\,\left({\nabla}_{\sigma}a_{\mu} + G_{\sigma\mu}\right) \right. \\ - \left. \frac{1}{16}\,\varepsilon_{abcd}\,{\omega}_{\rho}^{ab}\,
\left({\nabla}_{\sigma}{\omega}_{\mu}^{cd} + R_{\sigma\mu}^{cd}\right)\right],
\end{multline}
\begin{multline}\label{c4}
{\hat{e}}_{\mu}^{a} = {e}_{\mu}^{a} - \theta^{\rho \sigma}\left[a_{\rho}\left({\nabla}_{\sigma}{e}_{\mu}^{a} +
 \frac{1}{2}\,{\omega}_{\sigma}^{ab}\,{e}_{b \mu} + b_{\sigma}\,f_{\mu}^{a}\right)\right. \\ \left. +
\frac{1}{4}\,\varepsilon^{a}_{bcd}\,{\omega}_{\rho}^{bc}\left({\nabla}_{\sigma}f_{\mu}^{d} + \frac{1}{2}\,{\omega}_{\sigma}^{de}\,f_{e \mu} +
b_{\sigma}\,{e}_{\mu}^{d}\right)\right],
\end{multline}
\begin{multline}\label{c5}
{\hat{f}}_{\mu}^{a} = {f}_{\mu}^{a} - \theta^{\rho \sigma}\left[a_{\rho}\left({\nabla}_{\sigma}{f}_{\mu}^{a} +
\frac{1}{2}\,{\omega}_{\sigma}^{ab}\,{f}_{\mu}^{a} + \frac{1}{2}\,{\omega}_{\sigma}^{ab}\,{f}_{b \mu} + b_{\sigma}\,e_{\mu}^{a}\right)\right.  \\ \left. +
\frac{1}{4}\,\varepsilon^{a}_{bcd}\,{\omega}_{\rho}^{bc}\left({\nabla}_{\sigma}e_{\mu}^{d} + \frac{1}{2}\,{\omega}_{\sigma}^{de}\,e_{e \mu} +
b_{\sigma}\,{f}_{\mu}^{d}\right)\right],
\end{multline}
where $e_{b \mu} = \eta_{bc}\,e_{\mu}^{c}$ and $f_{b \mu} = \eta_{bc}\,f_{\mu}^{c}$.

\section{Appendix D}

If we write ${\hat{F}}_{\mu \nu}$ as
\begin{equation}\label{d1}
{\hat{F}}_{\mu \nu} = \frac{1}{4}\,{\hat{R}}_{\mu \nu}^{ab}\,\sigma_{ab} + {\hat{G}}_{\mu \nu}\,I + i\,{\hat{H}}_{\mu \nu}\,\gamma_{5},
\end{equation}
then we obtain from (\ref{efmunuone})
\begin{multline}\label{d2}
{\hat{R}}_{\mu \nu}^{ab} = {R}_{\mu \nu}^{ab} -  \frac{1}{2}\,\theta^{\rho \sigma}
\left[{\omega}_{\rho}^{ab}\left({\nabla}_{\sigma}G_{\mu \nu} + D_{\sigma}G_{\mu \nu}\right)
+  a_{\rho}\left({\nabla}_{\sigma}{R}_{\mu \nu}^{ab} + D_{\sigma}R_{\mu \nu}^{ab}\right)  \right.  \\  \left.   -
\frac{1}{2}\,{\omega}_{\rho}^{cd}\left({\nabla}_{\sigma}{H}_{\mu \nu} + {D}_{\sigma}{H}_{\mu \nu}\right)\varepsilon^{ab}_{cd}
- b_{\rho}\left({\nabla}_{\sigma}{R}_{\mu \nu}^{cd} + D_{\sigma}R_{\mu \nu}^{cd}\right)\varepsilon^{ab}_{cd}  \right.  \\  \left.   -
2\,{R}_{\mu \rho}^{ab}\, G_{\nu \sigma} + R_{\mu \rho}^{cd}\,{H}_{\nu \sigma}\,\varepsilon_{cd}^{ab} +
{H}_{\mu \rho}\,R_{\nu \sigma}^{cd}\,\varepsilon^{ab}_{cd}\right],
\end{multline}
\begin{multline}\label{d3}
{\hat{G}}_{\mu \nu} = {G}_{\mu \nu} -  \frac{1}{2}\,\theta^{\rho \sigma}\left[{a}_{\rho}\left({\nabla}_{\sigma}G_{\mu \nu} +
D_{\sigma}G_{\mu \nu}\right) - {G}_{\mu \rho}\,{G}_{\nu \sigma} \right.   \\
- b_{\rho}\left({\nabla}_{\sigma}{H}_{\mu \nu}^{ab} +
D_{\sigma}H_{\mu \nu}\right)
 +  H_{\mu \rho}\,H_{\nu \sigma} \\
\left. - \frac{1}{8}\,{\omega}_{\rho}^{ab}\left({\nabla}_{\sigma}{R}_{\mu \nu}^{cd} +
D_{\sigma}{R}_{\mu \nu}^{cd}\right)\eta_{ac}\,\eta_{bd}
+  \frac{1}{4}\,{R}_{\mu \rho}^{ab}\,{R}_{\nu \sigma}^{cd}\,\eta_{ac}\,\eta_{bd}\right],
\end{multline}
\begin{multline}\label{d4}
{\hat{H}}_{\mu \nu}  ={H}_{\mu \nu} -  \frac{1}{2}\,\theta^{\rho \sigma}\left[{a}_{\rho}\left({\nabla}_{\sigma}H_{\mu \nu} + D_{\sigma}H_{\mu \nu}\right) + b_{\rho}\left({\nabla}_{\sigma}{G}_{\mu \nu} + D_{\sigma}G_{\mu \nu}\right)  \right.   \\
- 2 {G}_{\mu \rho}\,H_{\nu \sigma}
 - {} 2 H_{\mu \rho}\,G_{\nu \sigma} -   \frac{1}{16}\,{\omega}_{\rho}^{ab}\left({\nabla}_{\sigma}{R}_{\mu \nu}^{cd} +
D_{\sigma}{R}_{\mu \nu}^{cd}\right)\varepsilon_{abcd}  \\
 \left. +  \frac{1}{8}\,{R}_{\mu \rho}^{ab}\,{R}_{\nu \sigma}^{cd}\,\varepsilon_{abcd}\right].
\end{multline}

\section{Appendix E}

If we expand the noncommutative quantities in (\ref{5.25}) as
\begin{equation}\label{e1}
{\hat{T}} = T + T^{(1)} + T^{(2)} + \cdots,
\end{equation}
\begin{equation}\label{e2}
{\hat{E}} = E + E^{(1)} + E^{(2)} + \cdots,
\end{equation}
\begin{equation}\label{e3}
{\hat{A}} =  A + A^{(1)} + A^{(2)} + \cdots,
\end{equation}
then we obtain in the first order in $\theta^{\mu \nu}$ the following results
\begin{multline}\label{e4}
{T_{\mu \nu}^{a}}^{(1)}   = \partial_{\mu}{e_{\nu}^{a}}^{(1)} -  \partial_{\nu}{e_{\mu}^{a}}^{(1)}
+ \left({\omega}_{\mu}^{ab}\,{e_{b\nu}}^{(1)} -  {\omega}_{\nu}^{ab}\,{e_{b\mu}}^{(1)}\right)  - 2 \left({f_{\mu}^{a}}^{(1)}\,b_{\nu}
- {f_{\nu}^{a}}^{(1)}\,b_{\mu}\right)  \\
+  \left({{\omega}_{\mu}^{ab}}^{(1)}\,e_{b\nu} -  {{\omega}_{\nu}^{ab}}^{(1)}\,e_{b\mu}\right)  - 2 \left(f_{\mu}^{a}\,b_{\nu}^{(1)}
- f_{\nu}^{a}\,b_{\mu}^{(1)}\right)    \\
+  \left[C_{1}\left(e_{\mu}^{a}, a_{\nu}\right) -  C_{1}\left(e_{\nu}^{a}, a_{\mu}\right)\right] - \epsilon^{a}_{\phantom{a}bcd}\, \left[C_{1}\left(f_{\mu}^{b}, \omega_{\nu}^{cd}\right) -  C_{1}\left(f_{\nu}^{b}, \omega_{\mu}^{cd}\right)\right],
\end{multline}
\begin{multline}\label{e5}
{U_{\mu \nu}^{a}}^{(1)} = \partial_{\mu}{f_{\nu}^{a}}^{(1)} -  \partial_{\nu}{f_{\mu}^{a}}^{(1)}
+ \left({\omega}_{\mu}^{ab}\,{f_{b\nu}}^{(1)} -  {\omega}_{\nu}^{ab}\,{f_{b\mu}}^{(1)}\right)  - 2 \left({e_{\mu}^{a}}^{(1)}\,b_{\nu}
- {e_{\nu}^{a}}^{(1)}\,b_{\mu}\right)  \\
+  \left({{\omega}_{\mu}^{ab}}^{(1)}\,f_{b\nu} -  {{\omega}_{\nu}^{ab}}^{(1)}\,f_{b\mu}\right)  - 2 \left(e_{\mu}^{a}\,b_{\nu}^{(1)}
- e_{\nu}^{a}\,b_{\mu}^{(1)}\right)  \\
+  \left[C_{1}\left(f_{\mu}^{a}, a_{\nu}\right) -  C_{1}\left(f_{\nu}^{a}, a_{\mu}\right)\right] - \frac{1}{4}\,\epsilon^{a}_{\phantom{a}bcd}\, \left[C_{1}\left(e_{\mu}^{b}, \omega_{\nu}^{cd}\right) -  C_{1}\left(e_{\nu}^{b}, \omega_{\mu}^{cd}\right)\right],
\end{multline}
where ${{\omega}_{\mu}^{ab}}^{(1)}, a_{\mu}, {e_{\mu}^{a}}^{(1)}$, ${f_{\mu}^{a}}^{(1)}$  are determined from the equations (\ref{c1}) - (\ref{c5}), and the bilinear operators $C_{1}( , )$ are given by expressions of the form (\ref{poisbrack}). For example, we have
\begin{equation}\label{e6}
C_{1}\left(e_{\mu}^{a} , a_{\nu}\right) = \theta^{\rho \sigma}\left({\nabla}_{\rho}e_{\mu}^{a}\,{\nabla}_{\sigma}a_{\nu}
- \frac{1}{2}{\tilde{R}}_{\sigma\mu\nu}^{\lambda}\,e_{\rho}^{a}\,a_{\lambda}\right),
\end{equation}
and similar expressions can be written for other bilinear operators in (\ref{e4}) and (\ref{e5}). In obtaining the expression (\ref{e6}) we used, in addition to (\ref{poisbrack}), the definitions (\ref{arrosig}).

\section{Appendix F}
Using the expressions (\ref{c1}) - (\ref{c5}), we obtain
\begin{equation}\label{f1}
{\omega}_{\mu}^{ab} = {\cal O}({\theta}^{2}),
\end{equation}
\begin{eqnarray}\label{f2}
{\hat{a}}_{\mu} = a_{\mu} - \frac{1}{2}\, \theta^{\rho\sigma}\,\left[a_{\rho}\left({\nabla}_{\sigma}a_{\mu} + G_{\sigma\mu}\right)
- b_{\rho}\left({\nabla}_{\sigma}b_{\mu} + H_{\sigma\mu}\right)\right] + O({\theta}^{2}),
\end{eqnarray}
\begin{eqnarray}\label{f3}
{\hat{b}}_{\mu} = b_{\mu} - \frac{1}{2}\, \theta^{\rho\sigma}\,\left[a_{\rho}\left({\nabla}_{\sigma}b_{\mu} + H_{\sigma\mu}\right)
+ b_{\rho}\left({\nabla}_{\sigma}a_{\mu} + G_{\sigma\mu}\right)\right] + {\cal O}({\theta}^{2}),
\end{eqnarray}
\begin{eqnarray}\label{f4}
{\hat{e}}_{\mu}^{a} = {e}_{\mu}^{a} -  \theta^{\rho\sigma}\,\left[a_{\rho}\left({\nabla}_{\sigma}{e}_{\mu}^{a}
+  b_{\sigma}\,{f}_{\mu}^{a}\right)\right] + {\cal O}({\theta}^{2}),
\end{eqnarray}
\begin{eqnarray}\label{f5}
{\hat{f}}_{\mu}^{a} = {f}_{\mu}^{a} -  \theta^{\rho\sigma}\,\left[a_{\rho}
\left({\nabla}_{\sigma}\,{f}_{\mu}^{a}
+  b_{\sigma}{e}_{\mu}^{a}\right)\right] + {\cal O}({\theta}^{2}).
\end{eqnarray}
Correspondingly, the first order noncommutative corrections of the strength tensor ${\hat{F}}_{\mu\nu}$  and torsion ${\hat{T}}_{\mu\nu}$ are
\begin{equation}\label{f6}
{\hat{R}}_{\mu\nu}^{ab} = {\cal O}({\theta}^{2}),
\end{equation}
\begin{eqnarray}\label{f7}
{\hat{G}}_{\mu\nu} = {G}_{\mu\nu} -  \frac{1}{2}\,\theta^{\rho\sigma}\,\left[a_{\rho}\left({\nabla}_{\sigma}{G}_{\mu\nu}
+  D_{\sigma}{G}_{\mu\nu}\right) - 2G_{\mu\rho}\,{G}_{\nu\sigma}  \right. \nonumber \\
- b_{\rho}\left({\nabla}_{\sigma}{H}_{\mu\nu} +  D_{\sigma}{H}_{\mu\nu}\right) + 2{H}_{\mu\rho}\,{H}_{\nu\sigma} +
{\cal O}({\theta}^{2}),
\end{eqnarray}
\begin{eqnarray}\label{f8}
{\hat{H}}_{\mu\nu} = {H}_{\mu\nu} -  \frac{1}{2}\,\theta^{\rho\sigma}\,\left[a_{\rho}\left({\nabla}_{\sigma}{H}_{\mu\nu}
+  D_{\sigma}{H}_{\mu\nu}\right) - 2{G}_{\mu\rho}\,{G}_{\nu\sigma}  \right.  \nonumber \\
+ b_{\rho}\left({\nabla}_{\sigma}\,{G}_{\mu\nu} +  D_{\sigma}{G}_{\mu\nu}\right) - 2{H}_{\mu\rho}\,{G}_{\nu\sigma}
+ {\cal O}({\theta}^{2}),
\end{eqnarray}
\begin{multline}\label{f9}
{\hat{T}}_{\mu\nu}^{a} = {\tilde{T}}_{\mu\nu}^{a} + \partial_{\mu}{e_{\nu}^{a}}^{(1)} -  \partial_{\nu}{e_{\mu}^{a}}^{(1)}
- 2\left({f_{\mu}^{a}}^{(1)}\,b_{\nu} -  {f_{\nu}^{a}}^{(1)}\,b_{\mu}\right)
- 2\left(f_{\mu}^{a}\,b_{\nu}^{(1)} -  f_{\nu}^{a}\,b_{\mu}^{(1)}\right)    \\
+  \left[C_{1}\left(e_{\mu}^{a}, a_{\nu}\right) -  C_{1}\left(e_{\nu}^{a}, a_{\mu}\right)\right]
+ {\cal O}({\theta}^{2}),
\end{multline}
\begin{multline}\label{f10}
{\hat{U}}_{\mu\nu}^{a} = {\tilde{U}}_{\mu\nu}^{a} + \partial_{\mu}{f_{\nu}^{a}}^{(1)} -  \partial_{\nu}{f_{\mu}^{a}}^{(1)}
- 2\left({e_{\mu}^{a}}^{(1)}\,b_{\nu} -  {e_{\nu}^{a}}^{(1)}\,b_{\mu}\right)
- 2\left(e_{\mu}^{a}\,b_{\nu}^{(1)} -  e_{\nu}^{a}\,b_{\mu}^{(1)}\right)   \\
+  \left[C_{1}\left(e_{\mu}^{a}, a_{\nu}\right) -  C_{1}\left(e_{\nu}^{a}, a_{\mu}\right)\right]
+  \left[C_{1}\left(f_{\mu}^{a}, a_{\nu}\right) -  C_{1}\left(f_{\nu}^{a}, a_{\mu}\right)\right]
+ {\cal O}({\theta}^{2}).
\end{multline}

\end{document}